\documentclass{aa}
\usepackage[varg]{txfonts}
\usepackage{graphicx}
\usepackage{t1enc}
\usepackage{epsfig}
\usepackage{rotating}
\usepackage[authoryear]{natbib}
\bibpunct{(}{)}{;}{a}{}{,}
\newcommand{\Tex}   {$T_\mathrm{ex}$}
\newcommand{\Trot}  {$T_\mathrm{rot}$}

\newcommand{\kms}   {km~s$^{-1}$}
\newcommand{\cmt}   {cm$^{-3}$}
\newcommand{\cmd}   {cm$^{-2}$}
\newcommand{\jpb}   {$\mathrm{Jy~beam^{-1}}$} 
\newcommand{\lo}    {$L_{\sun}$}
\newcommand{\mo}    {$M_{\sun}$}
\newcommand{\nh}    {NH$_3$}
\newcommand{\nth}   {N$_2$H$^+$}
\newcommand{\chtoh} {CH$_3$OH}
\newcommand{\hho}   {H$_2$O}
\newcommand{\hh}   {H$_2$}

\newcommand{\et}    {et al.}
\newcommand{\eg}    {e.\,g.,}
\newcommand{\ie}     {i.\,e.,}

\newcommand{\vel} {$v_\mathrm{LSR}$}
\newcommand{\velo} {$v$}

\newcommand{\hcop}  {HCO$^+$}
\newcommand{\htcop}  {H$^{13}$CO$^+$}
\newcommand{\tco}   {$^{13}$CO}
\newcommand{\cs}	{CS}

\newcommand{\nn}	{N$_2$}
\newcommand{\htp} {H$_{3}^{+}$}
\newcommand{\hii}	{\ion{H}{ii}}
\newcommand{\sfrs}	{star-forming regions}

\newcommand{\phnp}   {\phantom{0.}}

\newcommand{\phn}   {\phantom{0}}
\newcommand{\phnn}  {\phantom{0}\phantom{0}}

\newcommand{\chtcn}   {CH$_3$CN}

\begin{document}

	\title{N$_2$H$^+$ depletion in the massive protostellar cluster AFGL\,5142}
	
	\author{G. Busquet\inst{1} 
       	 \and
         R. Estalella\inst{1}
	\and
	Q. Zhang \inst{2}
	 \and
	S. Viti\inst{3}
	\and
	Aina Palau\inst{4} 
	\and
	P.~T.~P. Ho \inst{2,5}
	\and
  	\'A. S\'anchez-Monge\inst{1}
         }

  	\offprints{Gemma Busquet,\\ \email{gbusquet@am.ub.es}}

	\institute{Departament d'Astronomia i Meteorologia (IEEC-UB), Institut de 
	Ci\`encies del Cosmos, Universitat de Barcelona, Mart\'{\i} i Franqu\`es 1, E-08028 Barcelona, Spain
        \and
	 Harvard-Smithsonian Center for Astrophysics, Cambridge, MA, 02138
	 \and
	 Department of Physics and Astronomy, University College London, Grower Street,
	 London WC1E 6BT, UK
	 \and
	  Institut de Ci\`encies de l'Espai (CSIC-IEEC), Campus UAB,
         Facultat de Ci\`encies, Torre C-5 parell, E-08193 Bellaterra,
         Catalunya, Spain
	 \and
	 Academia Sinica Institute of Astronomy and Astrophysics, Taipei, Taiwan
	       }
 	 \date{Received / Accepted}

	  \authorrunning{G. Busquet \et}
 	 \titlerunning{N$_2$H$^+$ depletion in AFGL\,5142}

\abstract{}{We aim at investigating with high angular resolution the \nh/\nth\ abundance ratio toward
the high-mass star-forming region AFGL\,5142 in order to study whether the \nh/\nth\ ratio behaves similarly to the low-mass case, for which the ratio decreases from starless cores to cores associated with young stellar objects (YSOs).}{CARMA was used to observe the 3.2~mm continuum and \nth\,(1--0) emission toward AFGL\,5142. We used \nh\,(1,1) and (2,2), as well as \hcop\,(1--0) and \htcop\,(1--0) data available from the literature to study the chemical environment.
Additionally we performed a time-dependent chemical modeling of the region.}{The 3.2~mm continuum emission reveals a dust condensation of $\sim23$~\mo\ associated with the massive YSOs, deeply embedded in the strongest \nh\ core (hereafter central core). The dense gas
emission traced by  \nth\  reveals two main cores, the western core of $\sim0.08$~pc in size and the eastern core of
$\sim0.09$~pc, surrounded by a more extended and complex structure of $\sim0.5$~pc, mimicking the morphology of the \nh\ emission. The two cores are located to the west and to the east of the 3.2~mm dust condensation. Toward the central core the \nth\ emission drops significantly, indicating a clear chemical differentiation in the region. The \nth\ column density in the central core is one order of magnitude lower than in the western and eastern cores. Furthermore, we found low values of the \nh/\nth\ abundance ratio $\sim$50--100 toward the western and eastern cores, and high values up to 1000 associated with the central core. The chemical model used to explain the differences seen in the \nh/\nth\ ratio indicates that density, and in particular temperature, are key parameters in determining the abundances of both \nh\ and \nth.  The high density ($n\simeq10^{6}$~\cmt) and temperature ($T\simeq70$~K) reached in the central core allow molecules such as CO to evaporate from grain mantles. The CO desorption causes a significant destruction of \nth, which favors the formation of \hcop. This result is supported by our observations, which show that \nth\ and \hcop\ are anticorrelated in the central core. The observed values of the \nh/\nth\ ratio in the central core can be reproduced by our model for times $t\simeq4.5-5.3\times10^{5}$~yr while in the western and eastern cores  the \nh/\nth\ ratio can be reproduced by our model for times in the range $10^{4}-3\times10^6$~yr. }{The \nh/\nth\ abundance ratio in AFGL\,5142 does not follow the same trend as in regions of low-mass star formation mainly due to the high temperature reached in hot cores.}


\maketitle

\section{Introduction}

It is well known that N-bearing molecules, such as \nh\ and \nth, are excellent tracers of the interstellar dense gas because none
of these molecules deplete onto dust grains until densities reach  $\sim10^{6}$~\cmt\ 
\citep{bergin1997,tafalla2004,flower06}. Thus observations of these dense gas tracers have become a powerful tool to study
the sites of star formation. However, only a few observational studies have focused on the comparison of \nh\ and \nth\ cores, and these
studies, which were carried out toward low-mass star-forming regions, find that the \nh/\nth\  abundance ratio is around
60--90 close to the young stellar objects (YSOs), while it rises up to 140--190 in starless cores
\citep{caselli2002a,hotzel2004,friesen2010}. In the intermediate-mass cores surrounding the high-mass star IRAS~20293+3952, \citet{palau2007} find the same trend, with a
high \nh/\nth\ ratio, up to 300, in the cores with starless properties, and around 50 for the cores associated with YSOs,
clearly showing that chemical differentiation is important in the region. All these studies show that the \nh/\nth\ abundance
ratio is consistent with being a `chemical clock'.

%****INTRODUCTION TAKEN FROM THE PROPOSAL***** The fact that high-mass stars form associated with a cluster makes
%the study of the youngest massive stars a complex but intriguing task. Very young massive stars are found
%surrounded by multiple dense cores, and these cores can be in different evolutionary stages, already forming
%Young Stellar Objects (YSOs), or with no stellar activity associated (\eg\  \citealt{beuther2005,
%birkmann2006,felli2006,tej2006,davis2007,wang2008}). A very useful tool to study these cluster environments are
%the observations of dense-gas tracer molecules. In particular, nitrogen-bearing species, like \nh\ and \nth, are
%excellent tracers of dense gas because they do not deplete onto dust grains until densities of $\sim 10^6$~\cmt\
%are reached \citep{bergin1997,flower06}. Thus, one may consider that the main differences between these two
%molecules (once opacity effects have been taken into account through the treatment of their hyperfine structure)
%must arise because the cores formed at different times and/or because they formed with different initial
%conditions.

%very young massive stars
%However, the statistics of this ratio in high-mass \sfrs\ is very poor, and then one
%needs to further investigate the behavior of the \nh/\nth\ ratio in order to asses these
%initial findings.

%and chemical modeling

In order to investigate the behavior of the \nh/\nth\ abundance ratio in high-mass \sfrs, and see if  the \nh/\nth\ ratio in massive YSOs shows the same trend as in the low-mass regime, we
carried out observations and chemical modeling of the high-mass star-forming region AFGL\,5142, located at a distance of 1.8~kpc
\citep{snell1988} in the Perseus arm. The region was selected as good candidate to study the \nh/\nth\ ratio because the high angular resolution \nh\ emission reveals several dense cores, which have very different temperatures, and one of them is associated with star formation in clustered mode with the presence of hot cores. The region consists of two main centers of high-mass star formation in different
evolutionary stages. The brightest near-infrared source \citep{hunter1995} IRAS\,05274+3345, with a bolometric luminosity of
3.8$\times10^3$~\lo\ \citep{Carpenter1990}, lies near the western edge of the \nh\ gas \citep{estalella1993} and a lack of
dense gas emission  associated with it suggests that the IRAS source is a more evolved region, which is consistent with the
detection of optical nebulosity by \citet{eiroa1994}. Based on its position in the $J$ vs. $J-H$ diagram, \citet{chen2005}
classify the IRAS source as a B2 star of 3~Myrs. \citet{torrelles1992} detect
a faint radio continuum source, IRAS\,05274+3345 East (hereafter referred as AFGL~5142), about $30''$ to the east of the IRAS source, which coincides with the peak position
of the dust emission at 3.4~mm \citep{hunter1999} and the \nh\ emission peak \citep{estalella1993}. Assuming that the
centimeter emission arises from an optically thin \hii\ region, the flux density is equivalent to a zero-age main sequence
(ZAMS) star of spectral type B2 or earlier \citep{torrelles1992,hunter1995} with a luminosity of
$4\times10^3$~\lo. In the near-infrared, \citet{hunter1995} report a cluster of 28 embedded sources in a region of $1'$
($\sim$0.3~pc in radius) near the position of the radio continuum source. Recently, \citet{qiu2008} carried out Spitzer Space
Telescope IRAC and MIPS observations toward AFGL~5142. The authors identify 44 YSOs, 20 of them clustered around the central
massive star, and surrounded by a more extended and sparse distribution of young stars and protostars. In addition,
high angular resolution observations of maser emission reveal a cluster of \hho\ and \chtoh\ masers in an area of $\sim5''$
\citep{hunter1995,goddi2006,goddi2007}. Subsequent observations with the Very Large Array (VLA) in the A configuration at
8.4~GHz and with the Submillimter Array (SMA) at 1.2~mm with $\sim1''$ of angular resolution were carried out by \citet{zhang2007},
who find that the centimeter source is resolved into three peaks, and that the 3.4~mm source detected by \citet{hunter1999} actually
consists of five millimeter sources. Regarding the molecular outflow emission in this region, \citet{zhang2007}  identify three
molecular outflows in CO\,(2--1) and SO\,$6_{5}$--$5_{4}$, one of them coinciding with an \hcop\ outflow and the well-collimated
SiO jet detected by \citet{hunter1999}. All three outflows appear to originate from the dust condensation in a region of about
$3''$. All this information indicates that active star formation in clustered mode is taking place in AFGL\,5142.

The dense gas emission in this region has been studied with single-dish telescopes in \nh, \cs, HCN, \hcop, \chtoh, and \chtcn\ 
(\citealt{verdes-montenegro1989,estalella1993,hunter1995,hunter1999,cesaroni1999}). In particular, the dense gas emission
traced by the \nh\ molecule has been observed with high angular resolution using the VLA \citep{zhang2002}. The high angular
resolution \nh\ emission consists of a central and compact core associated with the dust condensation, harboring at least three
intermediate/high-mass young stars, surrounded by fainter \nh\ cores located in a more extended structure with no signs of
stellar activity  (no maser nor molecular outflow emission associated) indicating that the region may harbor cores in different
evolutionary stages. The presence of cores containing massive star(s) together with cores with no star-formation activity makes
this region a good choice to study how the \nh/\nth\ ratio behaves in high-mass \sfrs.

In this paper we report CARMA observations of the continuum emission at 3.2~mm and the dense gas traced by \nth\,(1--0) toward
AFGL\,5142. The paper layout is as follows: in \S\,2 we summarize our observations and the data reduction process. In \S\,3 we present
the main results for the continuum and \nth\ molecular line emission. In \S\,4 we analyze the molecular emission of
several species by computing their column density maps. In \S\,5 we show the main results of the chemical model to qualitatively reproduce the
abundances of the region. Finally, in \S\,6 we discuss our findings, and we list the main conclusions in \S\,7.

\section{Observations}

\begin{figure}[t]
\begin{center}
\begin{tabular}{c}
    \epsfig{file=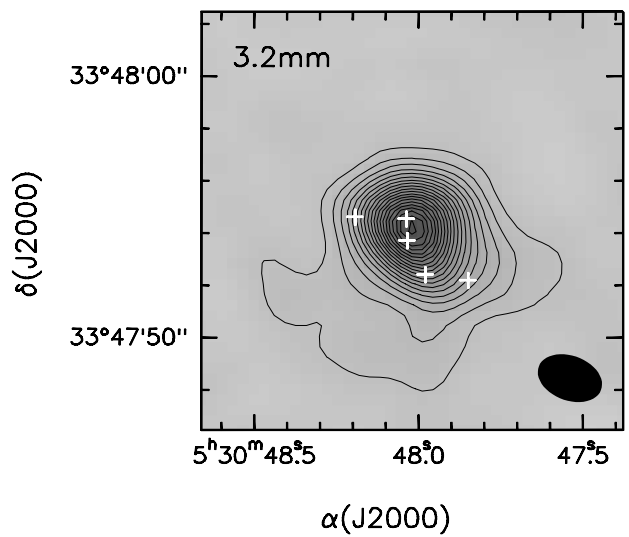,scale=0.8}\\
    \end{tabular}
     \caption{3.2~mm continuum emission toward AFGL~5142. Contour
     levels range from 3 to 60 $\sigma$ in steps of 3~$\sigma$, where $\sigma$ is the rms of the map, 0.7~m\jpb. The
     synthesized beam, shown in the bottom right corner, is $2\farcs53\times1\farcs71$,
     P.A.$=68.8\degr$. White crosses indicate the millimeter sources detected by
     \citet{zhang2007} at 1.2~mm with the SMA.}
\label{f3mmcont}
\end{center}
\end{figure}

\begin{table*}[t]{\center
\caption{Parameters of the hyperfine fits to the \nth\,(1--0) line for the central position of each core
}
\begin{tabular}{lcccccccc}
\hline\hline
&\multicolumn{2}{c}{Position$^{\mathrm{a}}$}
&$A\tau_{\mathrm{m}}^{\mathrm{b}}$
%&\sl{v}$^{\mathrm{c}}$
&\vel$^{\mathrm{c}}$
&$\Delta$\velo\
&&\Tex
&$N$(\nth)\\
\cline{2-3}
Core &  $\alpha$(J2000)& $\delta$(J2000) & (K~\kms) &(\kms) &(\kms) &$\tau_{\mathrm{m}}^{\mathrm{d}}$ &(K) & ($\times10^{12}$~\cmd) \\
\hline
central$^{\mathrm{e}}$ &05:30:48.02 &33:47:53.4 &0.51$\pm$0.04 &-0.97$\pm$0.06  &1.11$\pm$0.10  &0.10$\pm$0.03  &\phn8.2$\pm1.6$ & \phn2.8$\pm0.9$ \\  
eastern &05:30:48.73 &33:47:52.9 &1.35$\pm$0.02  &-2.92$\pm$0.01  &1.60$\pm$0.02  &0.10$\pm$0.02  &16.7$\pm1.1$ &18.5$\pm1.2$ \\
western &05:30:47.47 &33:47:51.7 &2.45$\pm$0.01  &-3.57$\pm$0.01  &1.31$\pm$0.03  &0.27$\pm$0.02  &12.4$\pm$0.8  &14.4$\pm1.2$ \\
\hline
\end{tabular}
\begin{list}{}{}
\textbf{Notes.} 
$^\mathrm{a}$ Positions, given in (h m s) and ($\degr$ $'$ $''$).
$^\mathrm{b}$ $A=f(J_{\nu}(T_{\mathrm{ex}})-J_{\nu}(T_{\mathrm{bg}}))$, where $f$ is the filling factor (assumed to be 1), \Tex is the excitation temperature of the
transition, $T_{\mathrm{bg}}$ is the background radiation temperature, and $J_{\nu}(T)$ is the intensity in units of temperature,
$J_{\nu}(T)=(h\nu/k)/(\exp(h\nu/kT)-1)$.
$^\mathrm{c}$ $v_\mathrm{LSR}$ corresponding to the $F_{1}F=01\to12$ hyperfine component . 
$^\mathrm{d}$ Optical depth of the main line, $F_{1}F=23\to12$, obtained from the fit.
$^\mathrm{e}$ Position of the central core taken from the \nh\ emission peak.
\end{list}
\label{hfspar}
}
\end{table*}

\begin{figure}[t]
\begin{tabular}{c}
    \epsfig{file=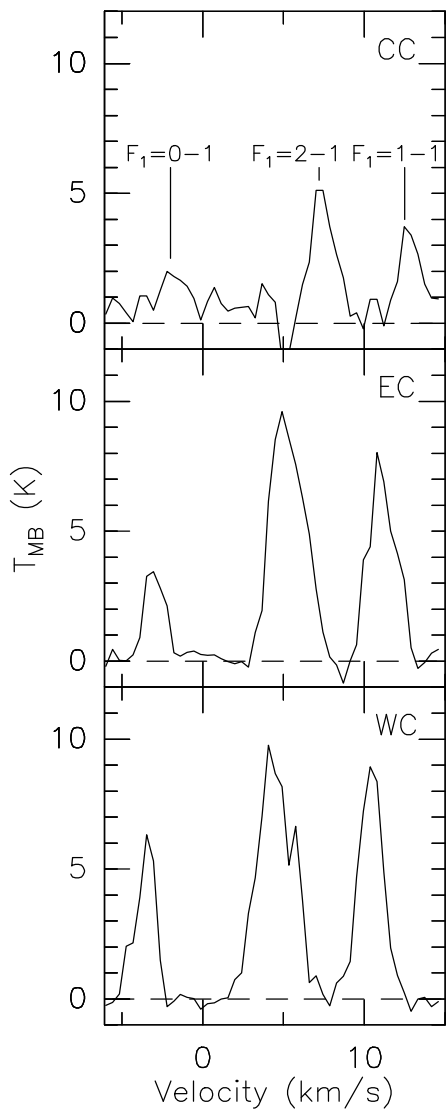,scale=1.2}\\
    \end{tabular}
     \caption{\nth\,(1--0) spectra toward three positions of AFGL\,5142. The three
     positions labeled on each panel are, from top to bottom, CC (\nh\ peak or central core), EC
     (peak of the eastern core), and WC (peak of the western core, see Fig.~\ref{nthpmom0}).}
\label{nthspec}
\end{figure}

\begin{figure*}[t]
\begin{center}
\begin{tabular}{c}
    \epsfig{file=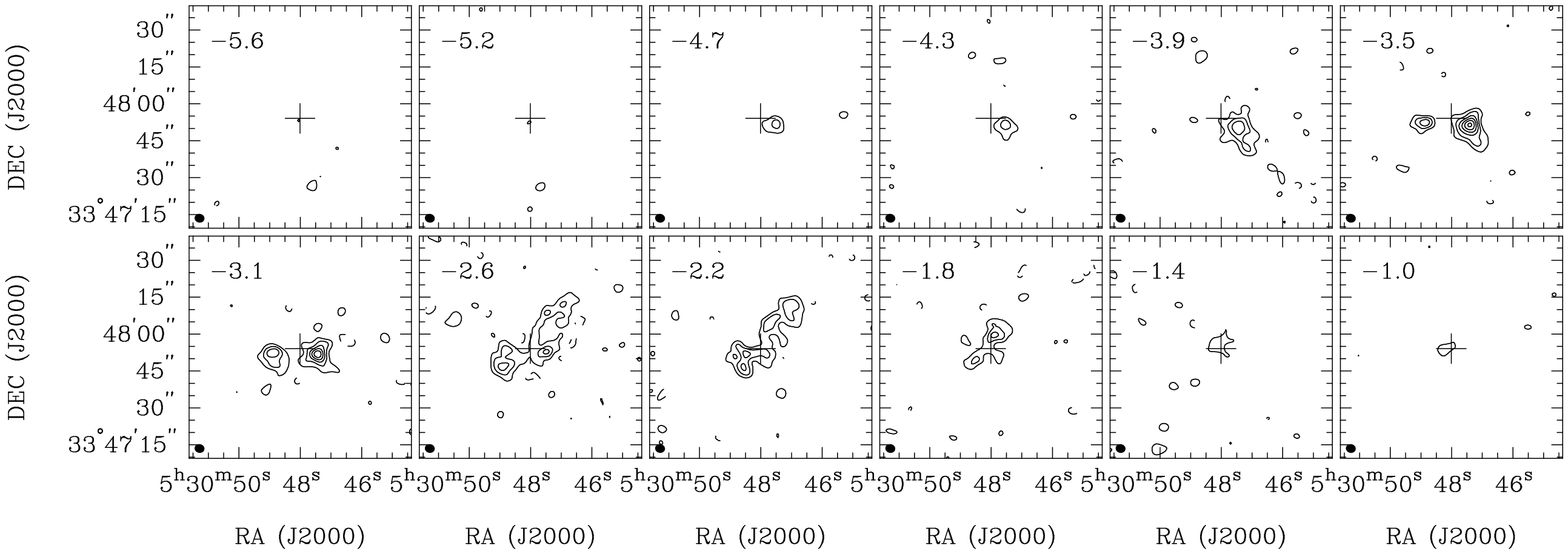,scale=0.60,angle=-90}\\
    \epsfig{file=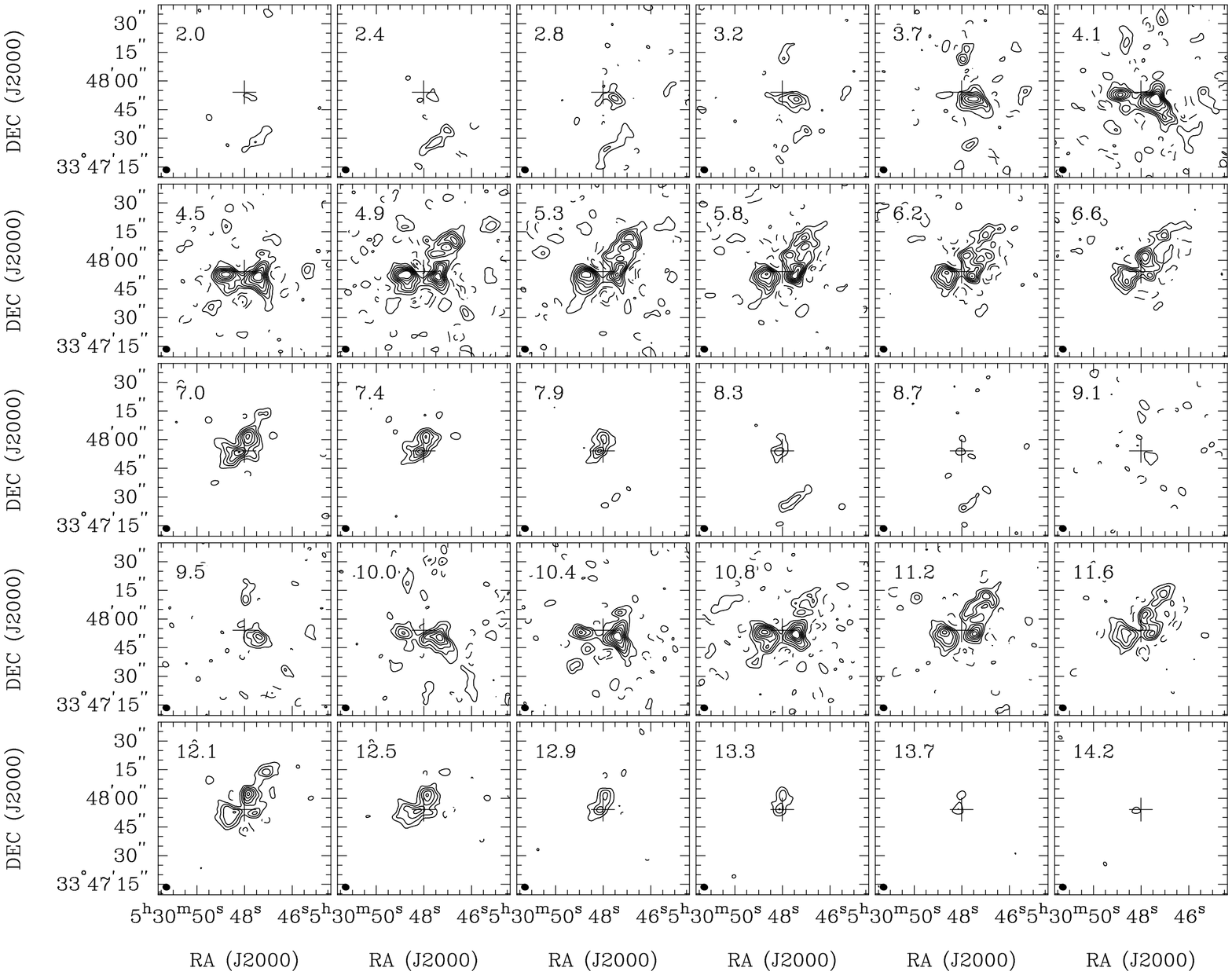,scale=0.656,angle=-90}\\
    \end{tabular}
     \caption{\emph{Top panel:} \nth\,(1--0) channel maps for the  hyperfine $F_{1}F=01\to12$ (isolated line) toward
     AFGL~5142. Contour levels are $-$2, 2, 4, 6, 8, 10, and 12 times the rms noise of the maps,
     0.055~\jpb. \emph{Bottom panel:} \nth\,(1--0) channel maps for the hyperfine $F_{1}F=21\to11$, $23\to12$, $22\to11$, and
     $F_{1}F=11\to10$, $12\to12$, and $10\to11$. Contour levels are $-$4, $-$2, 2, 4, 6, 8, 10, 12, and 14 times the rms of
     the map, 0.055~\jpb. In all panels the cross marks the position of the 3.2~mm continuum
     source. The synthesized beam is shown in the bottom left corner. Velocities are referred to the
     $F_{1}F=01\to12$ hyperfine, with the systemic velocity \vel=$-3$~\kms.}
\label{fn2hch}
\end{center}
\end{figure*}

The Combined Array for Research in Millimeter-wave Astronomy\footnote{Support for CARMA construction was derived from the
Gordon and Betty Moore Foundation, the Kenneth T. and Eileen L. Norris Foundation, the Associates of the California
Institute of Technology, the states of California, Illinois, and Maryland, and the National Science Foundation. Ongoing
CARMA development and operations are supported by the National Science Foundation under a cooperative agreement, and by
the CARMA partner universities.} (CARMA) was used to observe the 3.2~mm continuum and the \nth\,(1-0) emission toward
AFGL~5142. CARMA consists of six 10~m and nine 6~m antennas located at 2200 meters elevation at Cedar Flat in the Inyo
Mountains of California. The observations were carried out on 2007 February 4 and March 11 using the array in the C
configuration with 14 antennas in the array. The projected baselines ranged from 26 to 370~m. The phase center was set at
$\alpha=05^{\rm h}30^{\rm m}48\rm \fs02$, $\delta=+33\degr47\arcmin54\farcs47$. The FWHM of the primary beam at the
frequency of the observations was 132$''$ for the 6~m antennas and 77$''$ for the 10~m antennas. System temperatures were
around 250~K during both days.\\

The digital correlator was configured to observe simultaneously the continuum emission and the \nth\,(1--0) group of the hyperfine
transitions (93.176331~GHz, in the lower sideband). The continuum data were recorded in two $\sim$500~MHz bands covering the
frequency ranges 93.44--93.88~GHz and 96.74--97.18~GHz from the receiver lower and upper sidebands, respectively. We used two
consecutive bands of 8~MHz of bandwidth with 63 channels in each band, providing a spectral resolution of 0.42~\kms, to observe
the \nth\,(1--0) emission.\\

Phase calibration was performed with the quasars 0530+135 and 0555+398, with typical rms in the phases of 12$\degr$ and
8$\degr$, respectively. The absolute position accuracy was estimated to be around $0\farcs1$. Flux and bandpass
calibration was set by using 3C84. Data were calibrated and
imaged using the standard procedures in MIRIAD \citep{sault1995}. We combined the data from both days of observations. The
rms in the naturally weighted maps is $\sim0.7$~m\jpb\ in the continuum, and $\sim55$~m\jpb\ per $0.42$~\kms\ channel in
the line data. The synthesized beam of the continuum image is $2\farcs53\times1\farcs71$, P.A.$=68.8\degr$. For the line
emission, which is more extended than the continuum, we applied a $uv$-taper function of 52~k$\lambda$ ($3\farcs5$ in the
image plane) to improve the signal-to-noise ratio and to recover the extended emission. The resulting synthesized beam
is $4\farcs04\times3\farcs53$, with P.A.$=79.1\degr$. The continuum and \nth\ emissions were cleaned using a box around
the emitting region.

\section{Results}

\subsection{Continuum Emission}

Figure~\ref{f3mmcont} presents the continuum map at 3.2~mm. We  detected a single millimeter peak associated with IRAS\,05274+3345 East
(we could not resolve the two main 1.2~mm peaks MM-1 and MM-2, separated by about $1''$, detected with the SMA by \citealt{zhang2007}). The millimeter continuum emission has a single compact peak
component with a flux density of 107.1$\pm$1.6~mJy, surrounded by a more extended and flattened emission elongated toward
the west and south-east directions. A 2D Gaussian fit to the compact component yields a deconvolved size of
$2\farcs9\times2\farcs3$, P.A.$=16.5\degr$ (5200~AU at the distance of the source). The peak position is $\alpha(J2000)=05^{\rm
h}30^{\rm m}48\rm \fs01$; $\delta(J2000)=+33\degr47'54\farcs1$. The total flux density taking into account the extended emission
is 126.4$\pm$3.1~mJy, in very good agreement with the flux density at 88~GHz  reported by \citet{hunter1999}, which is
$\sim$125~mJy. Assuming that the dust emission is optically thin, and  using the dust opacity law of
\citet{hildebrand1983}, with a dust emissivity index of $\beta=1$ (see \citealt{zhang2007} for a description of its
derivation), and for a dust temperature of 45~K (based on the estimation of \citealt{hunter1999}) the total mass of dust
and gas from thermal continuum emission is 23~\mo\ (after correcting for the expected contribution, $\sim$4~\%, from the
ionized gas; see \citealt{hunter1999}). This mass is consistent with the mass estimated from the 1~mm SMA observations
($\sim21$~\mo). Although \citet{zhang2007} report a mass of 50~\mo, the difference arises from the value adopted for
the dust emissivity index, $\beta=$1.5 instead of 1. We note that there is a factor of 4 in the uncertainty of the mass,
mainly due to uncertainties in the dust opacity and the dust emissivity index.

\subsection{Molecular line emission: \nth}

\begin{figure}[!ht]
\begin{tabular}{c}
    \epsfig{file=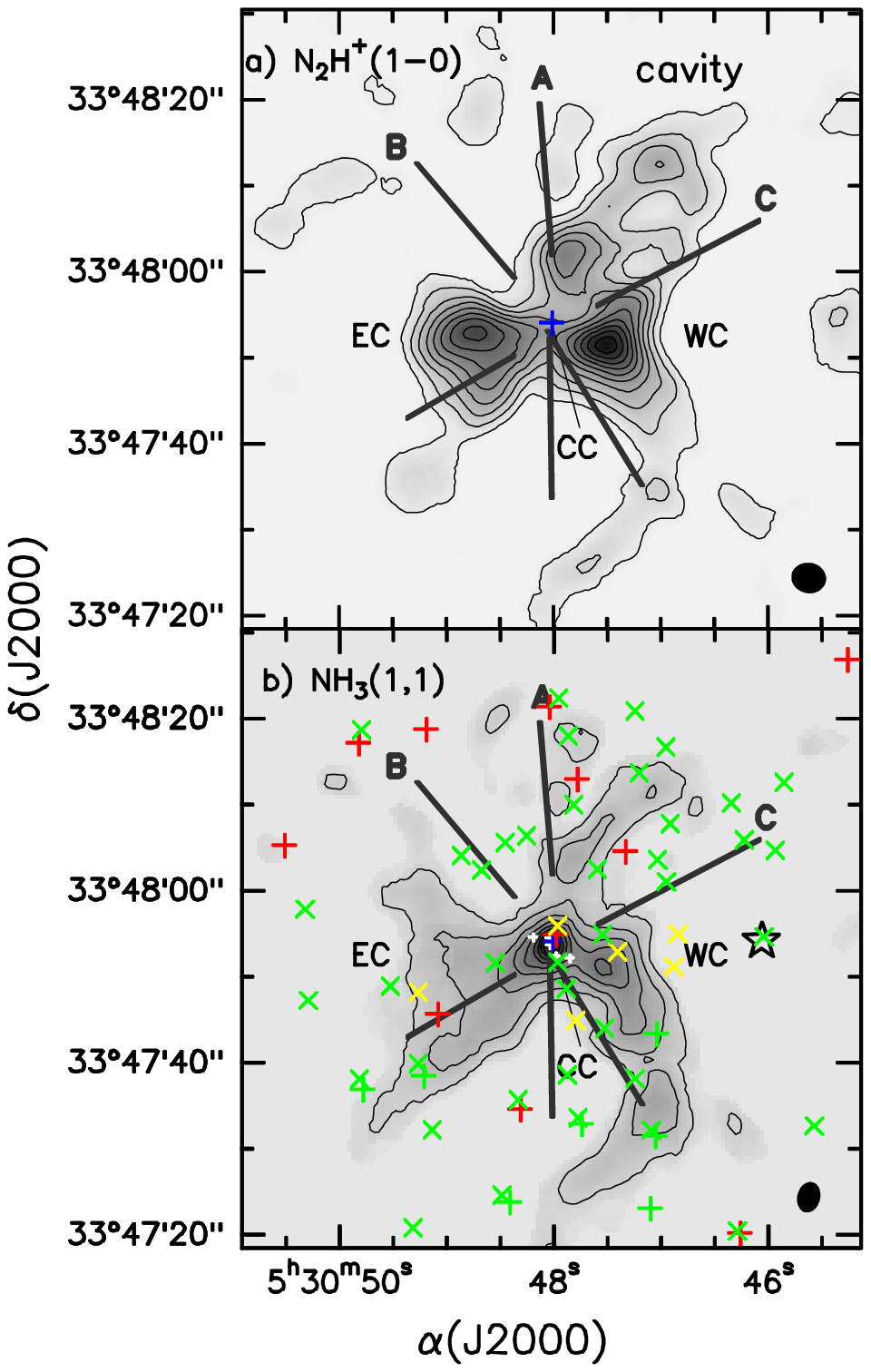,scale=0.8}\\
    \end{tabular}

     \caption{a) \nth\ zero-order moment integrated for all the hyperfine components of the (1--0) transition. Contours start at
     3~\%, increasing in steps of 10~\% of the peak intensity, 6.01~\jpb\kms. b) \nh\,(1,1) zero-order moment \citep{zhang2002}.
     Contours start at 10~\%, increasing in steps of 10~\% of the peak intensity, 0.202~\jpb\kms. The synthesized beams, $4\farcs04\times3\farcs53$ for \nth\ and $3\farcs58\times2\farcs64$ for \nh, are shown in
     the bottom right corner. In both panels the blue cross marks the position of the millimeter source reported in this work.
     Red crosses: type~I (protostars); green crosses: type~II (class II objects) from \citet{qiu2008}. Yellow and green tilted
     crosses are from \citet{chen2005} and represent Class~I luminous protostars with $M<\,5$~\mo, and Herbig Ae/Be stars or T Tauri
     stars, respectively. White crosses are the 5 millimeter peaks detected with the SMA by \citet{zhang2007}, and the star marks
     the position of the IRAS\,05274$+$3345. The western, eastern, and central cores are also labeled as WC, EC, and CC,
     respectively. The thick straight black lines represent the direction of outflows A, B, and C \citep{zhang2007}.}

\label{nthpmom0}
\end{figure}

\begin{figure}[t]
\begin{tabular}{c}
    \epsfig{file=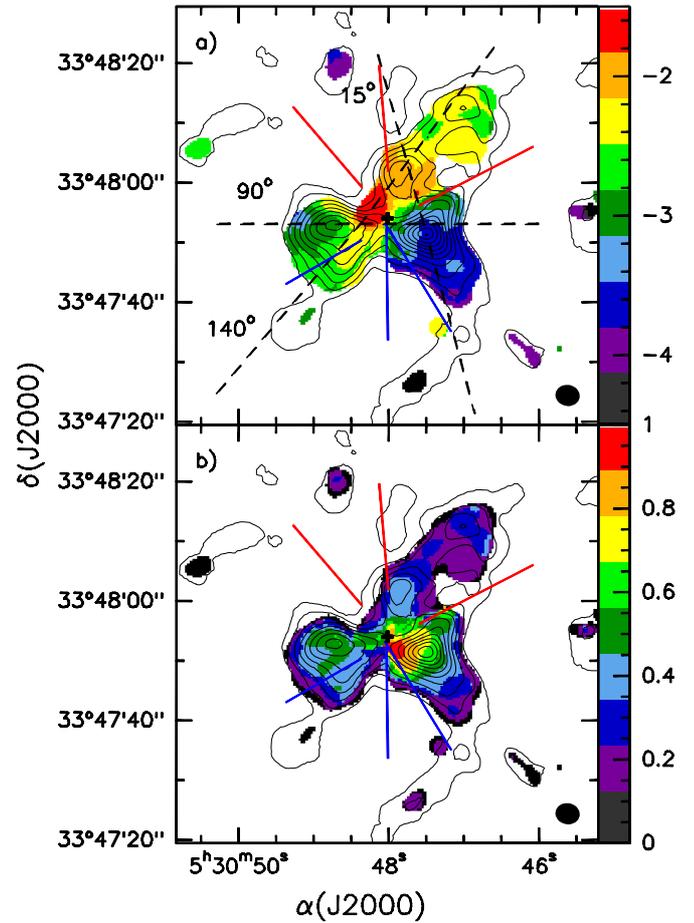,scale=0.8}\\
    \end{tabular}
     \caption{a) First-order moment map (velocity) for the hyperfine $F_{1}F=01\to12$
line of \nth\,(1--0) toward AFGL\,5142 (color scale). b) Second-order moment map (velocity dispersion) for
the hyperfine $F_{1}F=01\to12$ line of \nth\,(1-0) (color scale). In both figures black contours are the same as in
Fig~\ref{nthpmom0} (a), showing the \nth\,(1--0) emission, with contours starting at 3~\%,
and increasing in steps of 10~\% of the peak intensity. Color scales are in \kms. The
synthesized beam is shown in the bottom right corner of the image. The black cross marks the
position of the 3.2~mm source reported in this work. Blue and red lines represent the direction of outflows A, B, and C \citep{zhang2007}, and the black dashed line in the top panel indicate the position-velocity cuts (see Fig~\ref{pvplots}). Note that the second-order moment gives the velocity dispersion, and
must be multiplied by the factor 2$\sqrt{2\ln2}\simeq2.35$ to convert to full width to half maximum. }
\label{nthpmoms}
\end{figure}

In Fig.~\ref{nthspec} we show the \nth\,(1--0) spectra, not corrected for the primary beam response, at some selected positions of AFGL\,5142, and in Table~\ref{hfspar} we
present the line parameters obtained from the fit of the hyperfine spectra toward these positions. Figure~\ref{fn2hch} presents the velocity channel maps for the \nth\,(1--0) emission toward AFGL\,5142. The channel with
maximum intensity, for the hyperfine component $F_{1}F=01\to12$, was found at the velocity \velo$=-3.5$~\kms\ (the systemic velocity of the cloud is \vel$=-3$~\kms ).

%In Fig.~\ref{fn2hch} we show the velocity channel maps for the \nth\,(1--0) emission toward AFGL\,5142. The channel with
%maximum intensity, for the hyperfine component $F_{1}F=01\to12$, was found at the velocity of $v=-3.5$~\kms\  (the systemic velocity of the cloud is \vel$=-3$~\kms ). Fig.~\ref{nthspec} shows the
%\nth\,(1--0) spectra, not corrected for the primary beam response, at some selected positions, and in Table~\ref{hfspar} we
%present the line parameters obtained from the fit of the hyperfine spectra toward these positions. 

The zero-order moment map integrated for all the hyperfine transitions is presented in Fig.~\ref{nthpmom0}a. We also show (in 
Fig.~\ref{nthpmom0}b), for comparison, the \nh\,(1,1) integrated intensity map from \citet{zhang2002}
overlaid with the cluster of infrared sources reported in recent studies \citep{hunter1995,chen2005,qiu2008} in
order to have a complete view of the stellar content in AFGL\,5142. The overall structure of the  integrated \nth\,(1--0)
emission consists of two main cores, located to the west and to the east of the dust condensation (hereafter western and eastern
cores), surrounded by a more extended structure with an X-shape morphology, roughly following the same morphology as the \nh\
emission. As can be seen in Fig.~\ref{nthpmom0}, the western core seems to contain at least one infrared source embedded in the dense
gas, located close to the position of the \nh\ peak, and which displays typical characteristics of protostars \citep{chen2005},
whereas the infrared source lying close to the peak position of the eastern core has no associated infrared excess and has been
classified as a Herbig Ae/Be or T Tauri star \citep{chen2005}. However, there is no clear evidence that these infrared sources are really associated with the dense gas, and additionally, since the sensitivity of Spitzer observations is $\sim1$~\mo\ \citep{qiu2008}, and that of near-infrared studies is $\sim$0.2--0.4~\mo\ \citep{chen2005}, we cannot discard the possibility of more infrared sources embedded in the dense gas.
A 2D Gaussian fit  to the western and eastern cores yields deconvolved sizes of $\sim9\farcs3\times7\farcs6$
($\sim0.08\times0.07$~pc), and $\sim10\farcs4\times8\farcs3$ ($\sim0.09\times0.07$~pc), respectively. Interestingly, toward the
dust condensation where the intermediate/high-mass stars are deeply embedded in the strongest and compact \nh\ core
(hereafter central core), the \nth\ emission decreases drastically. However, it is important to note that the central core is
detected in \nth\ in some velocity channels, from 7~\kms\ to 8.7~\kms, whose peak positions are slightly shifted to the east
($\sim2''$) from the dust and \nh\ peaks.

We used the hyperfine component $F_{1}F=01\to12$ to study the kinematics of the \nth\ emission seen from Fig.~\ref{fn2hch} since this hyperfine component is strong and not blended with the other hyperfine components. In Fig.~\ref{nthpmoms}a we show the first-order moment map (intensity weighted mean
$v_{\mathrm{LSR}}$), and in Fig.~\ref{nthpmoms}b the second-order map (intensity weighted velocity dispersion) for this hyperfine.

The first-order moment map shows that the eastern and western cores appear at different velocities (see also the channel maps
shown in Fig.~\ref{fn2hch}). Toward the eastern core there is a small velocity gradient in the east-west direction, from
$\sim-3.5$~\kms\ to $\sim-1.8$~\kms. On the other hand, the western core shows a velocity gradient, from $\sim-4.7$~\kms\ to
$\sim-2.6$~\kms\ with increasing velocities from the southwest to the northeast. In fact, as can be seen in Fig.~\ref{nthpmoms}b the emission from the western core has very broad lines, with a velocity dispersion of $\sim0.8$--1~\kms, which corresponds for a Gaussian line profile, to
a full width at half maximum ($FWHM=2\sqrt{2\mathrm{ln}2}\times\sigma_{v}$, where $\sigma_{v}$ is the velocity dispersion) of $\sim1.8$--2.3~\kms\ (corrected for instrumental resolution) at the eastern side of the western core. This region of broad emission coincides with the passage of outflow~B of \citet{zhang2007}. The values around $\sim2$~\kms\ are
significantly higher than the thermal line width $\sim$0.2--0.3~\kms\ (estimated for a kinetic temperature of 25~K and 45~K,
respectively), indicative of \nth\ having a significant contribution from non-thermal processes, such as turbulence injected by
the molecular outflows, and/or global systematic motions. Regarding the eastern core, the typical value found for the velocity dispersion
is in the range $\sim0.3$--0.5~\kms, corresponding to line widths of $\sim0.2$--1.1~\kms\ (corrected for instrumental resolution).
In addition, there is a clumpy cavity structure toward the northwest clearly visible in the $-2.6$ and $-2.2$~\kms\ channel
maps of Fig.~\ref{fn2hch}, which appear redshifted by $\sim1$~\kms\ with respect to the systemic velocity.

\begin{figure}[t]
\begin{tabular}{c}
    \epsfig{file=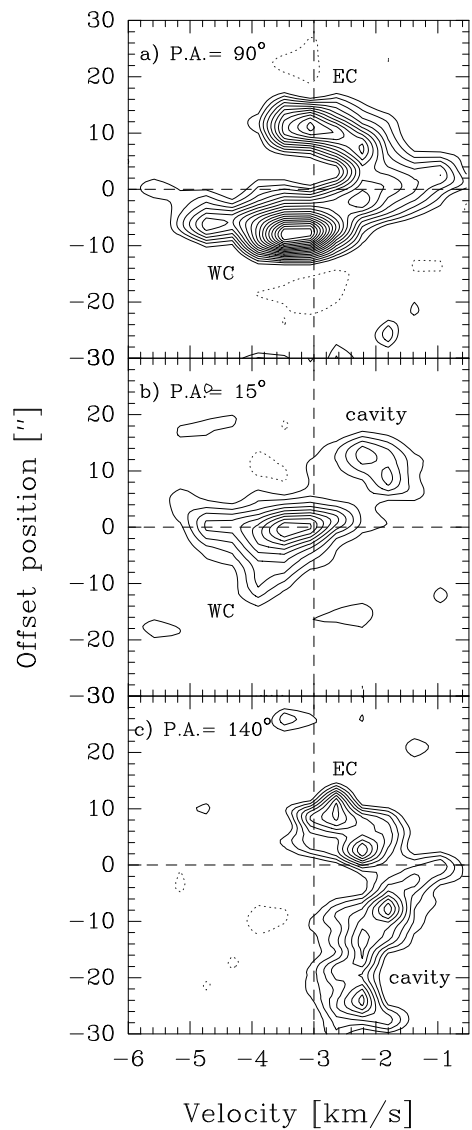,scale=0.6}\\
 
    \end{tabular}

	\caption{\nth\ p-v plot for the $F_{1}F=01\to12$ hyperfine at a)  PA$=$90$\degr$ (along the eastern and western
cores). Channel maps have been convolved with a beam $6''\times2''$, with P.~A. perpendicular to the direction of the cut. Contours start at
10~\%, increasing in steps of 5~\% of the peak emission, 0.408~\jpb.  The central position corresponds to the 3.2~mm source. Positive
offsets are toward the east. b) PA$=15\degr$. Channel maps have been convolved with a beam $2''\times2''$. Contours start at 10~\%, increasing in steps of 10~\% of the peak emission, 0.601~\jpb. The
central position is taken at the peak position of the western core, which is $\Delta\,x=-6.5''$, $\Delta\,y=-3''$  offset from the phase
center. Positive offsets are toward the northeast. c) PA$=140\degr$. Contours start at 20~\%, and increase in steps of
10~\% of the peak emission, 0.376~\jpb. Positive offsets are toward the southeast. The central position corresponds to $\Delta\,x=3.9''$,
$\Delta\,y=-0.59''$ offset from the phase center. In all panels, the vertical dashed line indicates the systemic velocity, \vel=$-3$~\kms. }

\label{pvplots}
\end{figure}

In Fig.~\ref{pvplots}a we present the position-velocity (PV) plot made across the east-west direction
(P.A.$=90\degr$), encompassing the \nth\ peaks of the western and eastern cores, which clearly shows a ring-like structure with
the western core having very broad line widths. The line broadening, of $\sim3.5$~\kms, is also seen in the PV-plot made across
the western core at P.A.$=15\degr$ (Fig.~\ref{pvplots}b). Additionally, we performed a third cut at
P.A.$=140\degr$ (Fig.~\ref{pvplots}c), picking up the eastern side of the cavity and the eastern core. From
this plot, it seems that there are two structures, a velocity gradient associated with the eastern core, of $\sim1.5$~\kms\ in a region
of 7$''$, and a curve structure associated with the cavity. Note that the curve structure seen in the PV-plot of Fig.~\ref{pvplots}c suggests that the cavity is expanding. Overall it seems that both cuts, at P.A.$=90\degr$ and P.A.$=140\degr$, reveal expanding motions, suggesting that these motions could be produced by a combination of stellar winds, radiation, and/or molecular outflows from the central cluster. Hence, the AFGL\,5142 cluster seems to be in process of disrupting the natal cloud. However, we will not further discuss the kinematics of the region since it is not the aim of this paper.

\section{Analysis}

\subsection{Column density maps}

\begin{figure}[t]
\begin{tabular}{c}
    \epsfig{file=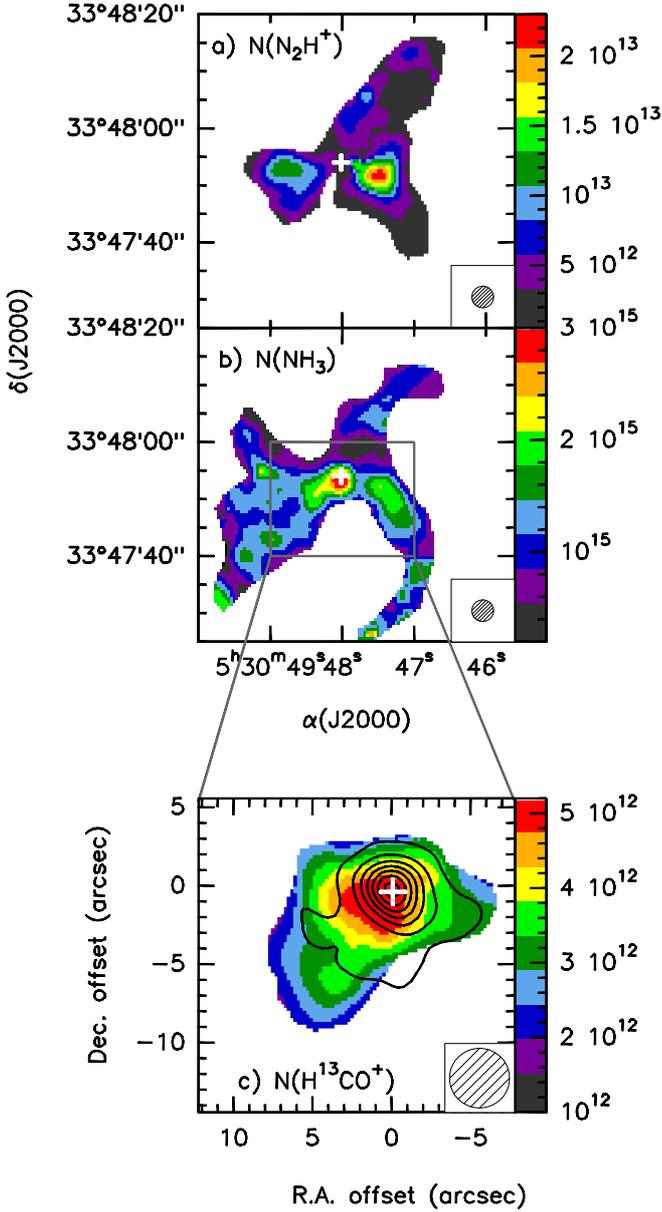,scale=0.8, angle=0}
    \end{tabular}
     \caption{a) \nth\ column density map. b) \nh\ column density map. c) \htcop\ column density map overlaid with the 3.2~mm continuum emission (contours). In all panels scale units are \cmd. The synthesized beam, $\sim4''$, is shown in the bottom right corner of each panel. The white
cross marks the position of the 3.2~mm peak.}
\label{fcoldens}
\end{figure}

We studied the chemical environment of AFGL\,5142 by analyzing the column density of several molecular species. We used the
\nth\,(1--0) (this work), \nh\,(1,1) and \nh\,(2,2)  (from \citealt{zhang2002}), and \hcop\,(1--0) and \htcop\,(1--0)  (from
\citealt{hunter1999}) data. In order to properly compare the emission of all the molecules, we convolved the \nth, \nh, \hcop,
and \htcop\ channel maps to obtain a final circular beam of $\sim4''$. We computed the column density maps by extracting the
spectra for positions in a grid of  $1''\times1''$. Using CLASS we fitted the hyperfine structure of each
spectrum for \nth\,(1--0) and \nh\,(1,1), and a single Gaussian for the \nh\,(2,2), \hcop\,(1--0) and \htcop\,(1--0). We
fitted only those positions with an intensity greater than 5$\sigma$ for \nth\,(1--0) and \nh\,(1,1) in order to ensure we are
detecting all the hyperfine components, whereas for the other molecular species we fitted the spectra with an intensity greater
than 4$\sigma$. In the following sections we show the method used to compute the column density for each molecule. The main
results are summarized in Table~\ref{coldens}.

\begin{figure}[t]
\begin{tabular}{c}
    \epsfig{file=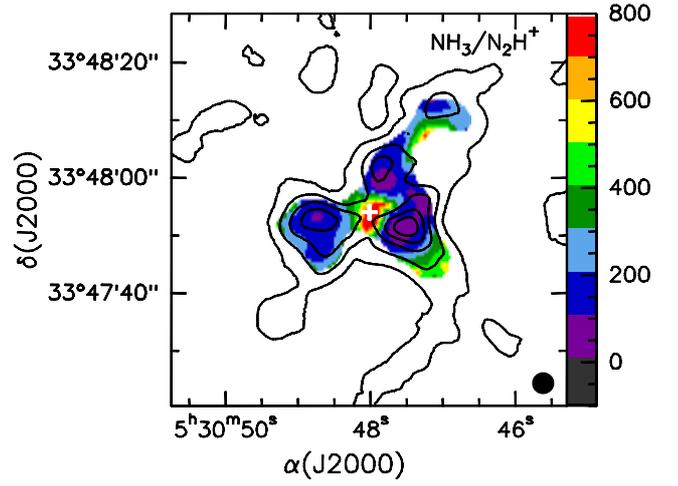,scale=0.75, angle=0}
    \end{tabular}

     \caption{Color scale: N(\nh)/N(\nth) abundance ratio map overlaid with the \nth\ emission (black contour), with contours starting at
3~\%, and increasing in steps of 20~\% of the peak intensity. The synthesized beam, $\sim4''$, is shown in the
bottom right corner. The white cross marks the position of the 3.2~mm peak.}

\label{fratio}
\end{figure}

\begin{table}[t]{\tiny
\caption{Average column density for each dense core
}
\begin{tabular}{lcccc}
\hline\hline
&$N$(\nh)  &$N$(\nth) &$N$(\htcop) &$N$(\nh)/$N$(\nth)   \\
Core&    (\cmd)  &(\cmd)  & (\cmd) &\\	    		    
\hline
central & $2.6\times10^{15}$   &$2.5\times10^{12}$	   &$4.5\times10^{12}$    &1000  \\
eastern & $1.4\times10^{15}$   &$1.1\times10^{13}$	   &\ldots		  &\phn130 \\
western & $1.8\times10^{15}$   &$1.8\times10^{13}$	   &\ldots		  &\phn100 \\
\hline
\end{tabular}
\label{coldens}
}
\end{table}

\subsubsection{\nth}

We used the hyperfine structure fitting program in CLASS \citep*{forveille1989} adopting the hyperfine frequencies given in
\citet{caselli1995} to determine the \vel, the intrinsic line widths, total optical depths, and excitation temperatures (\Tex)
in each position of the grid. The value of \Tex\ was derived assuming a filling factor of 1. As shown in Table~\ref{hfspar}, the
excitation temperature, \Tex, derived from the hyperfine fits in the eastern core, is in the range $\sim$14--16~K, while \Tex\ has
lower values, around $\sim$10--12~K and $\sim$8--9~K, for the western and central cores, respectively.

The results obtained from the fits indicate that the \nth\ emission is essentially optically thin for most of the region
($\tau_{\mathrm{TOT}}\simeq$0.3--0.6). However the optical depth in the western core reaches higher values, around
$\tau_{\mathrm{TOT}}\simeq$1.5--2. We calculated the \nth\ column density, corrected for the primary beam response, following the
expression given in \citet{caselli2002b}, and approximating the partition function to
$Q_{\mathrm{rot}}\simeq(kT_{\mathrm{ex}}/hB)\sim0.4473~T_{\mathrm{ex}}$, where $k$ is the Boltzmann constant, $T_{\mathrm{ex}}$
the excitation temperature, $h$ is the Planck constant, and $B$ is the rotational constant of \nth. Figure~\ref{fcoldens}a shows the
resulting \nth\ column density map. Clearly, we found important variation in the \nth\ column density in AFGL\,5142. The highest
value of the \nth\ column density, $3\times10^{13}$~\cmd, is reached in the western core. Toward the eastern core we found
values of the \nth\ column density around $\sim1\times10^{13}$~\cmd, while the \nth\ column density has the lowest values,
around $\sim$1--5$\times10^{12}$~\cmd, toward the millimeter condensation, \ie\ toward the central core (see Table~\ref{coldens}). The values found in the western and eastern cores are in good
agreement with the \nth\ column densities reported by \citet{Pirogov2003,Pirogov2007}, and \citet{Fontani2006} for dense molecular cloud
cores with massive stars and star clusters, and for a sample of high-mass protostellar candidates, respectively. In addition,
the values of the \nth\ column density found in the western and eastern cores of AFGL\,5142 are also consistent with those reported in recent studies
conducted toward massive star-forming regions observed with interferometers (\eg\ \citealt{palau2007,beuther2009}), but those obtained in the central core are clearly below. It
is important to emphasize the difference of 1 order of magnitude in the \nth\ column density in the central core  compared to the western and eastern cores. The uncertainty in the \nth\ column density is of the order of 25--50~\%, and it has been estimated taking into account the uncertainty in the output parameters of CLASS and the uncertainty in the calibration.

\subsubsection{\nh}

The optical depth of the main line derived from the \nh\ emission is in the range $\tau_{\mathrm{TOT}}\simeq$1.5--2.5 toward the central and compact
\nh\ core containing the intermediate/high-mass stars, while the emission in the extended structure is optically thin.

From the results of the fits of the \nh\,(1,1) and \nh\,(2,2) spectra we derived the rotational temperature
(\Trot) and computed the \nh\ column density map following the procedures described in \citet{ho1983} and \citet{harju1993}
(see also the appendix of \citealt{Busquet2009} for a 
description of the method). It is worth noting that the rotational temperature map obtained from
this analysis is similar to the \nh\,(2,2)/\nh\,(1,1) intensity map presented in \citet{zhang2002}.

%by assuming that the excitation temperature and the intrinsic line width are the same for both \nh\,(1,1) and \nh\,(2,2), and
%following the standard procedures (\citealt{ho1983,harju1993,sepulveda1993,anglada1995}). In addition, we assumed that only
%metastable levels up to level $(3,3)$ are populated and that \Trot\ is the same for each pair of rotational levels

%which has been obtained from the intensity ratio of the $(2,2)$ and the $(1,1)$ lines. 

The \nh\ column density map, corrected for the primary beam attenuation, is shown in Fig.~\ref{fcoldens}b. The
map shows small variations along the entire cloud. Contrary to \nth, the \nh\
column density reaches the maximum value, around $3\times10^{15}$~\cmd, toward the central core, at the
position of the millimeter condensation, and where we found the minimum \nth\ column density (see Table~\ref{coldens}). The values found for the \nh\ column density toward AFGL\,5142 are
similar to those obtained by \citet{palau2007} in the massive star forming region IRAS\,20293+3952. The uncertainty in the \nh\ column density is of the order of $\sim$15--25~\%, estimated from the uncertainty in the rotational temperature  (see \citealt{Busquet2009}), the output parameters of CLASS, and the uncertainty in the calibration.

%This analysis assumes implicitly that the physical conditions of the gas are homogeneous along the
%line-of-sight.

\subsubsection{\htcop}

To derive the \htcop\ column density, we assumed that \htcop\,(1--0) emission is optically thin while
\hcop\,(1--0) is optically thick. The \htcop\,(1--0) optical depth has been estimated from the ratio of the
\htcop\ and \hcop\ integrated intensity emission obtained from Gaussian fits and excluding the
high-velocity components of \hcop\,(1--0) coming from molecular outflow emission. 

Assuming that all levels are populated according to the same excitation temperature \Tex, and approximating the partition function to $k$\Tex/$hB$,  with B being the rotational constant of the molecule, the column density of a linear molecule obtained from the transition $J\to~J-1$ is:

\begin{equation}
N=\frac{3k}{4\pi^{3}}\frac{1}{\mu^{2}\nu_{10}J}T_\mathrm{ex}\frac{\exp{\big(\frac{J(J+1)}{2}\frac{h\nu_{10}}{kT_\mathrm{ex}}\big)}}{\exp{\big(\frac{J~h\nu_{10}}{kT_\mathrm{ex}}\big)}-1}\tau_{0}\Delta\,v,
\label{eqcoldens}
\end{equation}
where $\mu$ is the electric dipole momentum of the molecule, $\nu_{10}$ the frequency of the transition $J=1\to0$, and $\tau_{0}$ is the opacity at the line center.

For the particular case of \htcop\,(1--0) molecular transition, $\mu=3.88$~Debye$=3.88\times10^{-18}$~(cgs), $\nu_{10}=86.754$~GHz, the column density can be written, in practical units, as

\begin{equation}
\Big[\frac{N(\mathrm{H^{13}CO^{+}})}{\mathrm{cm^{-2}}}\Big]=2.56\times10^{11}\frac{1}{1-e^{-4.16/T_\mathrm{ex}}}\Big[\frac{T_\mathrm{ex}}{\mathrm{K}}\Big]\tau_{0}\Big[\frac{\Delta\,v}{\mathrm{km/s}}\Big].
\label{eqhtcopcoldens}
\end{equation}

In Fig.~\ref{fcoldens}c we
present the \htcop\ column density map toward the central core of AFGL\,5142. An obvious feature is that the
\htcop\ column density map is correlated with the \nh\ column density map, reaching the maximum value, around
$5\times10^{12}$~\cmd\ toward the millimeter condensation, but is anticorrelated with the \nth\ column density.
The values obtained for the \htcop\ column density are similar to those found toward regions of high-mass star
formation (\eg\ \citealt{Zinchenko2009}). Taking into account the uncertainty in the output parameters of CLASS and the uncertainty in the calibration, we estimated 
an uncertainty in the \htcop\ column density around $\sim$30--45~\%.

\subsection{The \nh/\nth\ abundance ratio map}

Figure~\ref{fratio} shows the \nh/\nth\ abundance ratio map toward AFGL\,5142. First of all, it is important to remark that
the \nth\ spectra toward some positions of the central core have low signal-to-noise, so only a few positions with high enough
signal-to-noise ratio were fitted properly, and hence the values obtained toward the central core, specially those toward the
south western side of the millimeter condensation, are overestimated due to the interpolation process. A clear feature seen from the
\nh/\nth\ ratio map is that the highest values, ranging from $\sim$400 up to 1000, are reached close to the position of the
central core (\ie\ harboring a cluster of millimeter sources). We explicitly checked that such a high values are real, not produced by the interpolation process. Toward the western core, which could contain a low-mass YSO, we
found values of the \nh/\nth\ ratio around 50--100, whereas in the eastern core, maybe containing a more evolved object, the \nh/\nth\
ratio is around 50--100 toward the center of the core and increases up to 200 at the edges. In Table~\ref{coldens} we give the average value of
the \nh/\nth\ ratio for these three cores. The uncertainty estimated for the \nh/\nth\
abundance ratio is around $\sim$30--60~\%. As already pointed out by \citet{zhang2002}, the VLA \nh\ data recover 65~\% of the flux observed by the Effelsberg 100~m telescope \citep{estalella1993}, and the missing flux arises mainly from smooth structures with scales larger than 50$''$. Unfortunately, we can not estimate the amount of flux filtered out by CARMA in \nth\ due to the lack of short spacing information. However, since the CARMA \nth\ and VLA \nh\  images are sensitive to similar angular scales, the \nh/\nth\ ratio is not seriously affected by differences in the missing flux.

\section{Chemical modeling}

\subsection{The UCL\_CHEM model} 

We employed UCL\_CHEM, a time- and depth- dependent chemical model similar to that used in \citet{viti99,viti01,viti04} to study the behavior of the \nh/\nth\ abundance ratio. For simplicity, in this
study we performed time-dependent single-point calculations. The model consists of a two-phase calculation: in phase~I we follow the
chemical and dynamical evolution of a collapsing core. This phase starts from a fairly diffuse medium ($n\simeq400$~\cmt) in neutral atomic
form (apart from a fraction of hydrogen in \hh) that undergoes a collapse following the free-fall collapse law described in
\citet{Rawlings1992} until densities typical of massive dense cores ($n\simeq10^{5}-10^{7}$~\cmt) are reached. During this time,
atoms and molecules from the gas freeze on to the dust grains and they hydrogenate where possible, as in \citet{viti99}.
The degree of freeze out or depletion, defined in terms of the depletion of the CO molecule, is a free parameter (although linked with density, see Rawlings et al. 1992), which we explore in this study.
During phase~I we adopted a constant temperature of 12~K, and we assumed a standard value for the cosmic ionization rate,
$\zeta=3\times10^{-17}$~s$^{-1}$. The initial gas-phase elemental abundances relative to hydrogen nuclei
adopted in this work, based on the findings of \citet{sofia01}, are listed in Table~\ref{iniabundances}. We included 127
gas-phase species and 42 surface species interacting in 1869 chemical reactions adapted from the UMIST~2006 database \citep{Woodall2007}. 
During the collapse
phase (phase~I) we let the chemistry develop for 3~Myrs, regardless of when the final density is reached.
% \textbf{We would like to note that, for convenience, we will refer as intial conditions those achieved at the end of phase~I.}

\begin{table}[t]{\center
\caption{Initial gas-phase elemental abundances relative to hydrogen nuclei.
\label{tjcmt}}
%\begin{center}
%\begin{scriptsize}
\begin{tabular}{ll}
\hline\hline
Atom
&Abundance\\
\hline
H	&1.0\\
He	&0.075\\
O	&4.45$\times10^{-4}$\\
C	&1.79$\times10^{-4}$\\
N	&8.52$\times10^{-5}$\\
S	&1.43$\times10^{-6}$\\
Mg	&5.12$\times10^{-6}$\\
\hline
\end{tabular}

%\end{scriptsize}
\label{iniabundances}
}
%\end{center}
\end{table}

\begin{table}[t]{\center
\caption{Parameters of the UCL\_CHEM model}
\begin{tabular}{lcc}
\hline\hline
Parameter &CC$^{\mathrm{a}}$ &WC and EC$^{\mathrm{b}}$ \\
\hline
Depletion$^{\mathrm{c}}$ (\%) &99 & 99 \\
Gas Density$^{\mathrm{c}}$ (\cmt) &10$^6$ &10$^5$ \\
$A_{\mathrm{v}}^{\mathrm{c}}$ &41 &10 \\
T$_{\mathrm{max}}^{\mathrm{d}}$ &70 &25 \\
\hline
\end{tabular}
\begin{list}{}{} \textbf{Notes.} $^\mathrm{a}$ CC: central core. $^\mathrm{b}$  WC and EC: western and eastern cores. $^\mathrm{c}$ At the end of phase~I. $^\mathrm{d}$ Maximum value reached during phase~II.
\end{list}
\label{modelpar}
}
\end{table}

%Model
%&Depletion$^{\mathrm{a}}$
%&Gas Density$^{\mathrm{a}}$
%&$A_{\mathrm{v}}^{\mathrm{a}}$
%&T$_{\mathrm{max}}^{\mathrm{b}}$
%\\
%&(\%)
%&(\cmt)
%&&(K)
%\\
%\hline
%Central Core\\
%\hline
%A  &99 &$10^5$ &\phn\phn6 &70    \\
%B  &56 &$10^5$ &\phn\phn6 &70   \\
%C  &33 &$10^5$ &\phn\phn6 &70      \\
%D   &99 &$10^6$ &\phn41 &70   \\
%E  &51 &$10^6$ &\phn41 &70  \\
%F  &30 &$10^6$ &\phn41 &70     \\  
%G  &99 &$10^7$ &388 &70   \\
%H  &48 &$10^7$ &388 &70  \\
%I  &28 &$10^7$ &388 &70     \\      
%\hline
%WC and EC\\
%\hline
%J  &99 &$10^5$ &\phn10 &25   \\
%K  &57 &$10^5$ &\phn10 &25  \\
%L  &34 &$10^5$ &\phn10 &25     \\
%M   &99 &$10^6$ &\phn79 &25   \\
%N  &51 &$10^6$ &\phn79 &25  \\
%O  &30 &$10^6$ &\phn79 &25     \\  
%P   &99 &$10^7$ &773 &25   \\
%Q  &48 &$10^7$ &773 &25  \\
%R  &28 &$10^7$ &773 &25    

In phase~II we explored the chemical evolution of the core harboring a newly
born massive star by simulating the presence of an infrared source at the center of the core, and by subjecting the core to a
time-dependent increase of the gas and dust temperature (assumed to be the same because of the high densities considered), which was performed differently depending on the core we were
modeling. For the central core we adopted a maximum temperature of $\sim70$~K, whereas the temperature toward the western and eastern cores is around $\sim25$~K (both derived from  \nh\ data, see \citealt{zhang2002}). During phase~II we performed the same
treatment for the increasing temperature and evaporation from grains as in  \citet{viti04}, which includes the experimental
results on desorption from grains and the evaporation of icy mantles formed in star-forming regions
\citep{collings2003a,collings2003b,collings2004}. In this case, the evaporation of a fraction of mantle species occur when the
temperature for a particular desorption event is reached (see \citealt{viti04} for a complete description). In phase~II we
followed the evolution of each core for 3~Myr. 

% primera versio
%Table~\ref{modelpar} lists the parameter choices for each model computed.

%\begin{table}[b]{
%\caption{Standard model parameters used in all of our models
%\label{tjcmt}}
%\begin{tabular}{lc}
%\hline
%\hline
%Grain Albedo				&0.5\\
%Grain size				&0.1~$\mu$m?\\
%Gas-to-dust mass ratio  		&100\\
%Cosmic ray ionization rate		&3$\times10^{-17}$~s$^{-1}$\\
%Mean photon scattering by grain 	&0.8\\
%External FUV radiation intensity	&1 Habing$^{\mathrm{a}}$\\
%\hh\ formation rate coeddificien	&$3\times10^{-18}\sqrt{\mathrm{T}}exp(-\mathrm{T}/1000)$~\cmt\,s$^{-1}$?\\
%\hline
%\end{tabular}
%\begin{list}{}{}
%\item[$^\mathrm{a}$] The standard Interstellar Radiation Field intensity is $I_{\odot}=1.6\times10^{-3}$~erg\,\cmd\,s$^{-1}$
%\citet{Habing1968} 
%\end{list}
%\label{standardpar}
%}
%\end{table}

%Although the eastern core does not contain any embedded source, we
%also run phase~II to allow an increase of the temperature, up to $\sim25$~K, that could be due to an external
%heating from the nearby molecular outflows. 

\subsection{Results}

%\begin{figure}[t]
%\begin{tabular}{c}
%    \epsfig{file=plots/phaseI_ratio_cc1.ps,scale=0.39, angle=0}\\
%    \end{tabular}
 %    \caption{\nh/\nth\ abundance ratio as a function of time during phase~I for the central core and for densities in the range of $n\simeq10^{5}$-$10^7$~\cmt.} 
%\label{fphaseI}
%\end{figure}

\begin{figure*}[t]
\begin{tabular}{cc}
    \epsfig{file=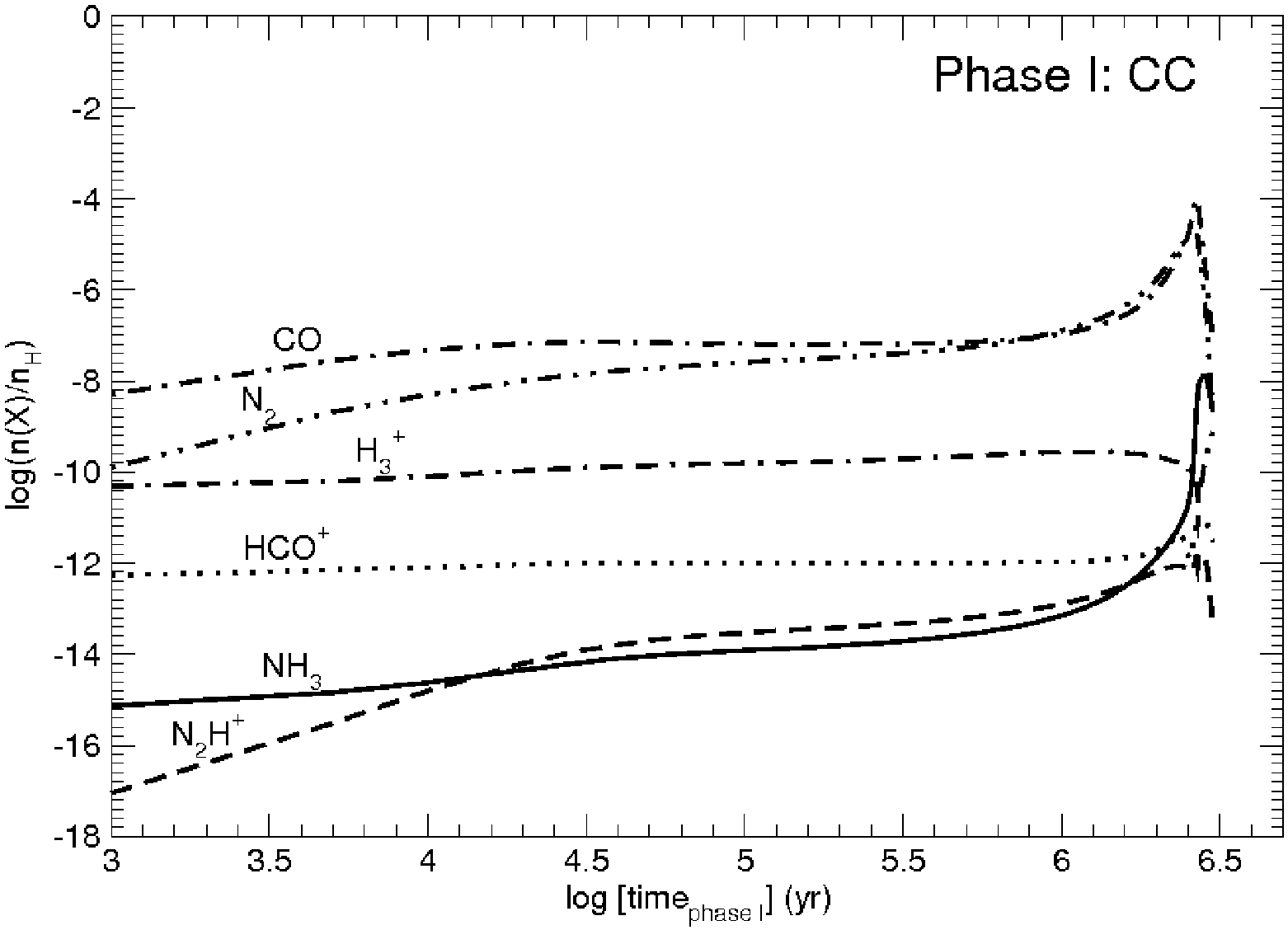,scale=0.37, angle=0}&
    \epsfig{file=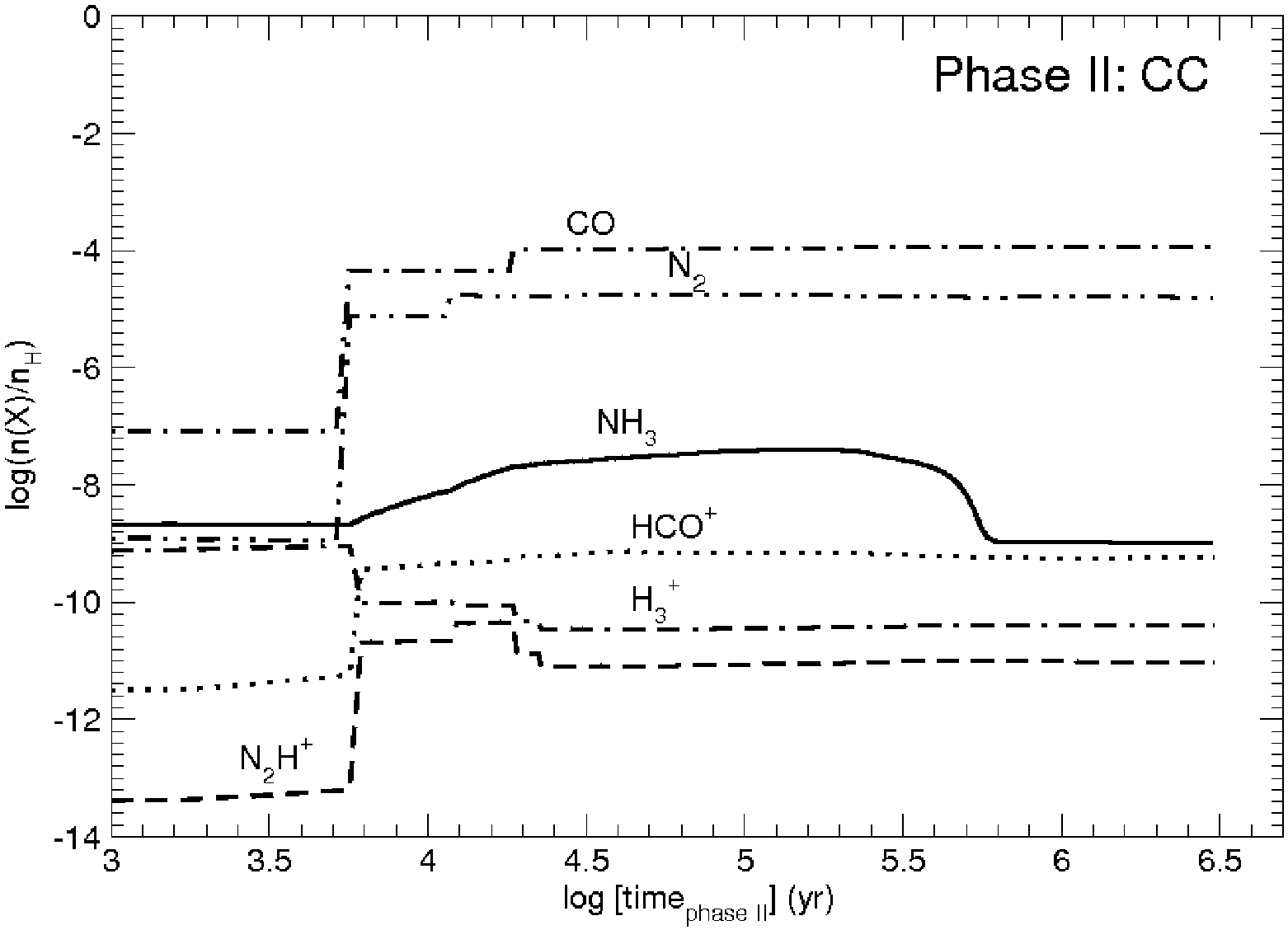,scale=0.37, angle=0}\\
    \epsfig{file=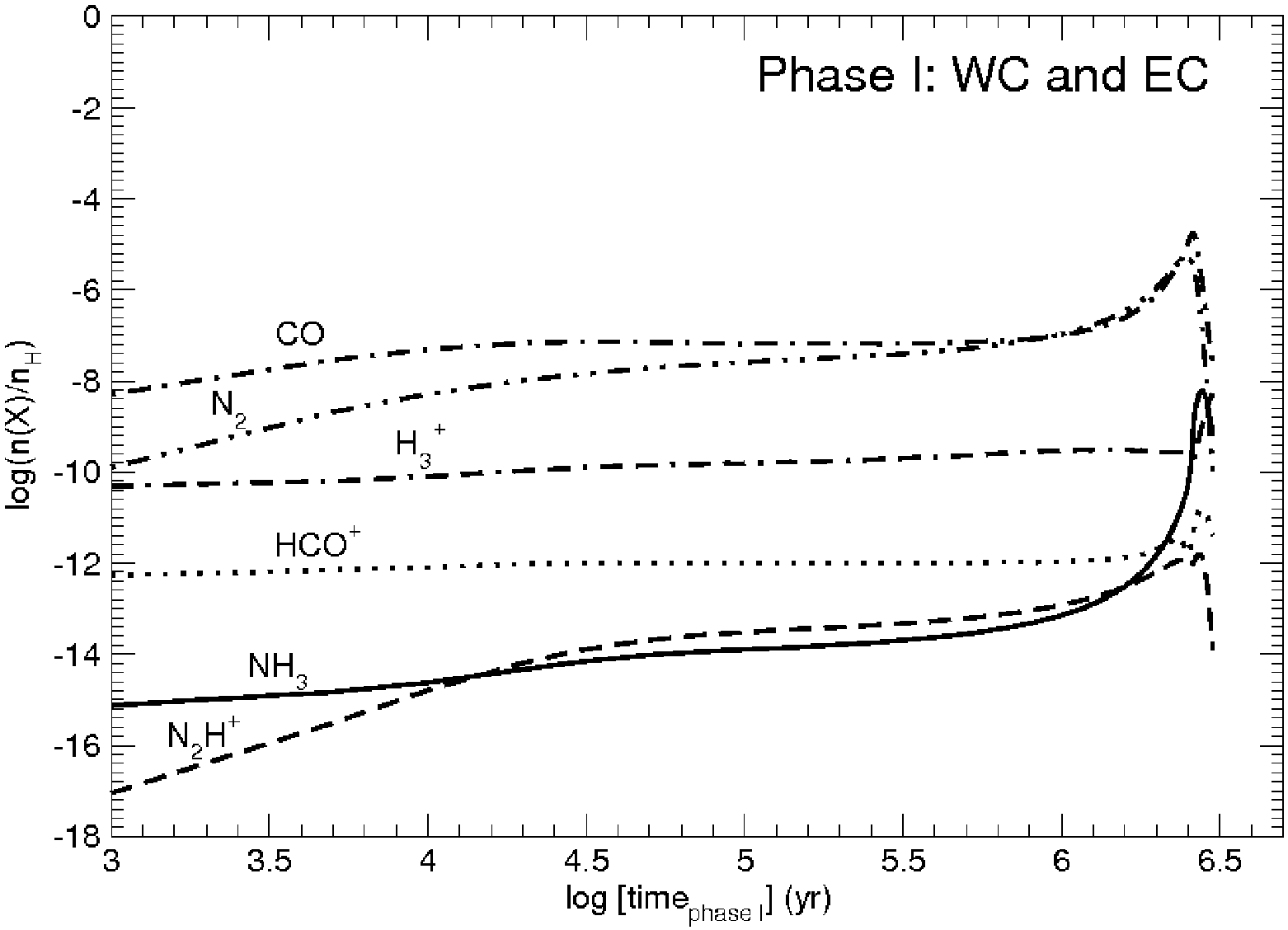,scale=0.38, angle=0}&
    \epsfig{file=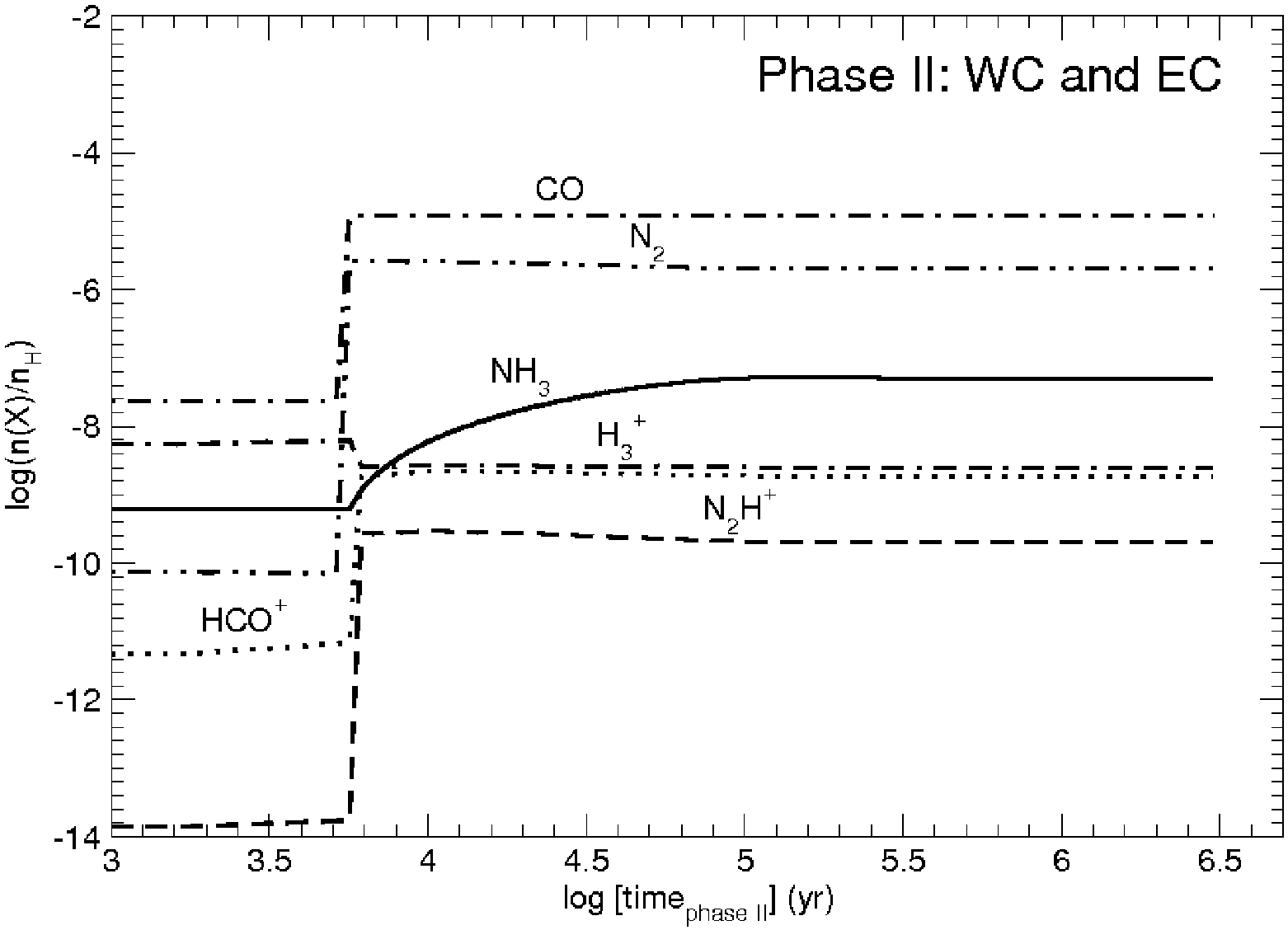,scale=0.38, angle=0}\\
     \epsfig{file=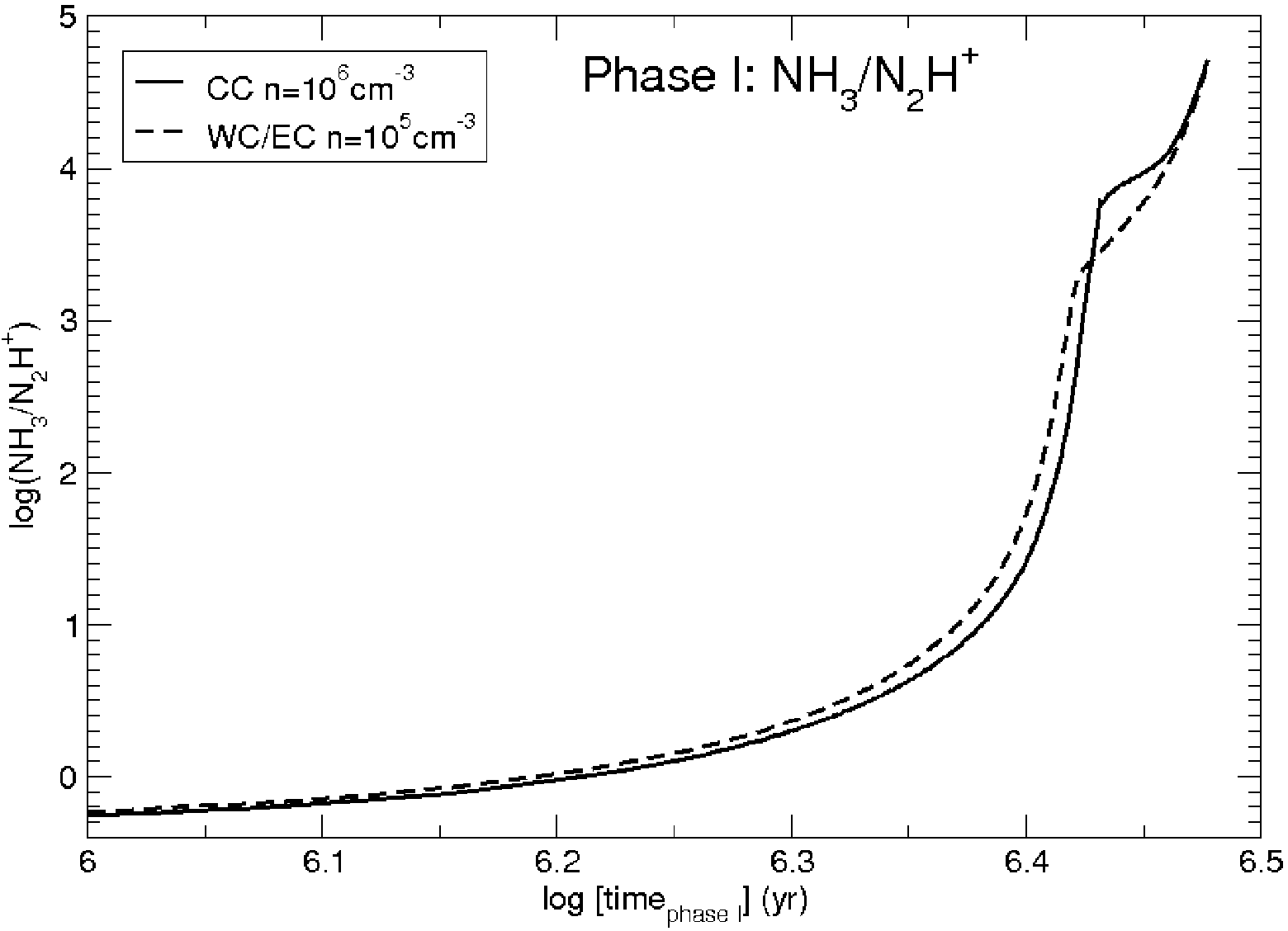,scale=0.38, angle=0}&
     \epsfig{file=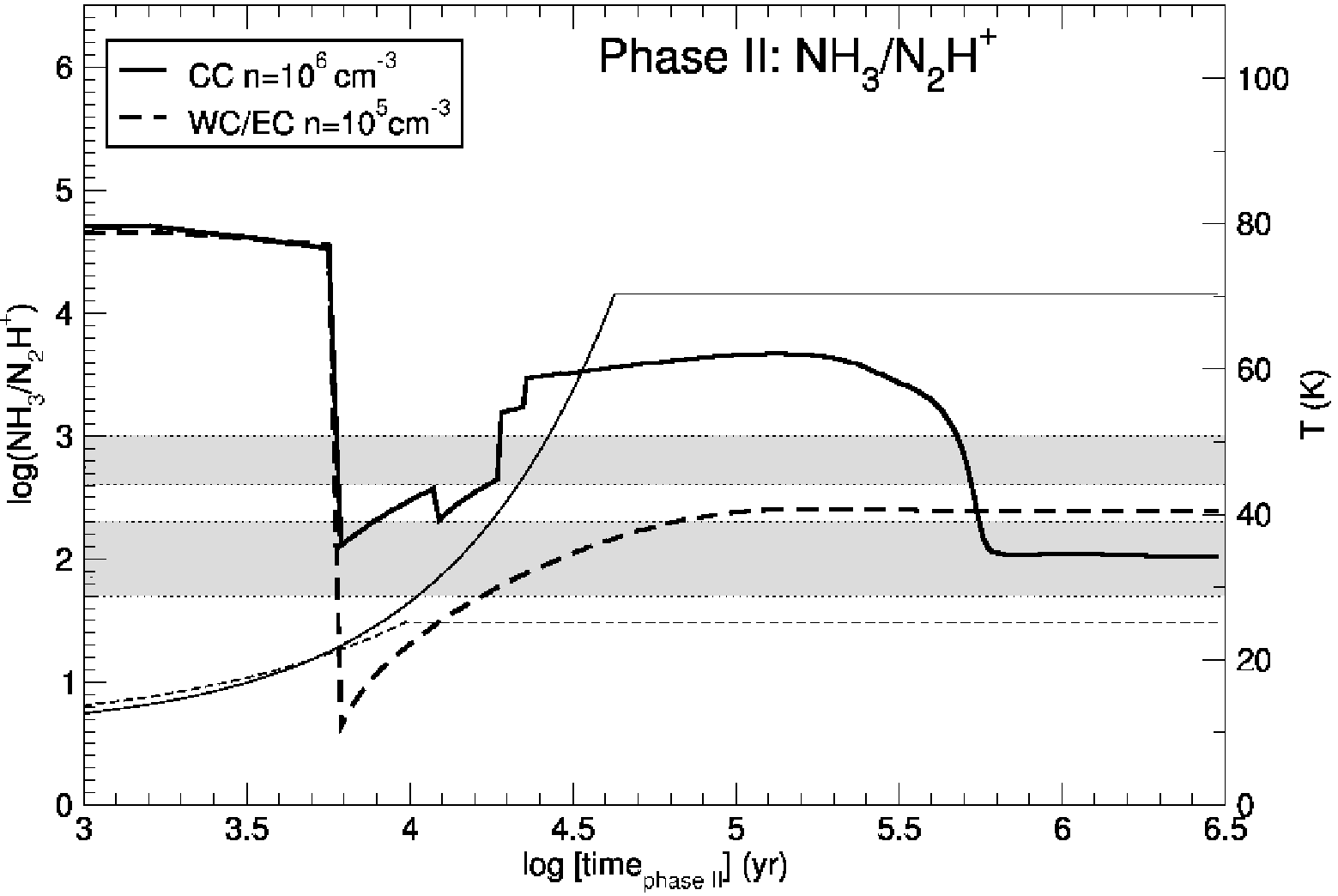,scale=0.38, angle=0}\\
    \end{tabular}

     \caption{\emph{Left:} Fractional abundances with respect to hydrogen as a function of time during phase~I for the central core (top panel) and the western/eastern cores (middle panel). The  \nh/\nth\ abundance ratio for both the central core and the western/eastern core are also shown in the bottom panel. \emph{Right:}  Same as left panels but for phase~II calculation. Note that the y-axis range according to the model. In the bottom right panel we also show the temperature as a function of time, indicated by the thin lines. The two shaded areas mark the range of values (400--1000 for the central core and 50--200 for the western/eastern cores) of the \nh/\nth\ abundance ratio estimated from the observational data.}
\label{modelfigs}
\end{figure*}

%\nh/\nth\ abundance ratio for the central core for a density in the range of $n\simeq10^{5}$--$10^{7}$~\cmt, and using
%different percentages of depletion at the end of phase~I. \emph{Top panel:} high depletion ($\sim99~\%$). \emph{Middle panel:}
%moderate depletion ($\sim50~\%$). \emph{Bottom panel:} low depletion ($\sim30~\%$). \emph{\textbf{Right:}} Same as left panels but for
%the western and eastern cores. Note that the y-axis range according to the model. In the top panels we also show the temperature as a function of time, indicated by the %thin line. The two thin dashed horizontal lines shown in the top panels mark the range of values of the \nh/\nth\ abundance ratio estimated from the observational data. }

We computed a grid of models to investigate the differences observed in the \nth\ column density and the \nh/\nth\ abundance
ratio. For the central core we followed the
chemical evolution of a core of 0.02~pc in diameter, whereas for the eastern
and western cores we adopted a size of 0.04~pc. We explored a range of densities, $n\simeq10^{5}$--10$^{7}$~\cmt, and for a
particular density we used different percentages of freeze-out at the end of phase~I. From this analysis we found that the observed values of the \nh/\nth\ ratio can be reproduced by our model using a high degree of depletion at the end of the collapse phase (phase~I), and by adopting a density $n\simeq10^6$~\cmt\ for the central core and $n\simeq10^5$~\cmt\ for the western/eastern cores. It is worth noting that the model for the central core  is supported by our estimate of the density from the dust continuum emission. Adopting the mass derived in this work (see Sect.~3.1) and a size of $6''$, we obtained a density
of $\sim5\times10^6$~\cmt, consistent with the density adopted in the chemical model. In Table~\ref{modelpar} we list the parameter choices for the two selected models (CC and WC/EC models), which can reproduce both the observed \nh\ and \nth\ column densities and the \nh/\nth\ abundance ratio. In the following sections, we discuss the main results obtained by our chemical model for phase~I and phase~II.

%In Fig.~\ref{fphaseI} we show the \nh/\nth\ abundance ratio  for the central core for densities in the range $n\simeq10^{5}-10^{7}$~\cmt\  as a function of time for phase~I %calculation  of some selected models listed in Table~\ref{modelpar}. We note that for the western and eastern cores the only difference is in the size of the core, and the \nh/%\nth\ abundance ratio during this phase is extremely similar to that of the central core. During phase~I  the \nh/\nth\ abundance ratio behaves in the same way for the range of %densities considered, for both the central core and the western and eastern cores, and only during the later steps it depends on the adopted density. In all cases, the \nh/\nth\ %ratio starts from low values and it reaches values around 1000 and even higher for $t>2.5\times10^6$~yr. 
% SV:

%This suggests that  long-lived starless cores  are associated with high values of the \nh/\nth\ abundance ratio. In the following sections we present the main results of our chemical model obtained for phase~II calculation, after the protostar %is born.

\subsubsection{Fractional abundances and the \nh/\nth\ abundance ratio during phase~I}

\begin{table*}[t]
\begin{center}
\caption{Results obtained from the chemical model
\label{tjcmt}}
\begin{tabular}{lcccc}
\hline\hline
\\
&
time
&$N$(\nh)
&$N$(\nth)
&\nh/\nth\
\\
Model
&(yr)
&($\times10^{14}$~\cmd)
&($\times10^{11}$~\cmd)
&
\\
\hline
%A &$2.3\times10^{4}$--2.9$\times10^4$ &2.00--1.30 &3.00--3.20 &\phn700--400  \\
%B &$1.0\times10^{3}$--5.7$\times10^3$  &0.24--0.26 &0.32--0.54         &\phn700--480  \\
%C &$1.0\times10^{3}$--5.7$\times10^3$ &0.20--0.22 &0.36--0.58 &\phn560--400 \\
CC &$1.9\times10^{4}$ &\phnp17                  &\phnp38 &450        \\
CC &$4.5\times10^{5}$--5.3$\times10^5$  &\phnp11--3.20 &0.80 &1400--400  \\
WC \& EC &$1\times10^{4}$-3$\times10^{6 }$ &2.80--10 &56--41 &\phnn50--240  \\
\hline
\end{tabular}
\label{tableres}
\end{center}
\end{table*}

%\begin{figure*}[t]
%\begin{tabular}{cc}
%  \epsfig{file=plots/phaseII_modelD_nh3.n2hp.hcop.ps,scale=0.39, angle=0}&
%    \epsfig{file=plots/phaseII_modelJ.nh3.n2hp.hcop.ps,scale=0.39, angle=0}\\
%    \end{tabular}
%     \caption{Fractional abundances with respect to hydrogen as a function of time for the central core (left panel) and the western and eastern
%     cores (right panel). The vertical dashed lines indicate the region for which the \nh/\nth\ ratio agree with the observational estimation. In the right panel only one line is %indicated becasue the \nh/\nth\ ratio agrees until the last time step.} 
%\label{modelabundances}
%\end{figure*}

In Fig.~\ref{modelfigs}~(left panels) we present the fractional abundances (with respect to hydrogen nuclei) of some molecular species of interest (\ie\ CO, N$_2$, H$_{3}^{+}$, \nh, \nth, and \hcop) together with the \nh/\nth\ abundance ratio as a function of time for the central core (CC model) and the western/eastern cores (WC/EC model) during phase~I. First of all, we note that the fractional abundances obtained for the central core (top panel) and the western/eastern cores (middle panel) do not show significant differences because they only differ in the core size and the final density reached at the end of this phase. 
As can be seen in Fig.~\ref{modelfigs}, the fractional abundance of CO, \nn, \nh, \nth, and \hcop\ increases with time (with a more pronounced increase for the case of CO, \nn, and \nh), reaching its maximum value at times $t\simeq2.5\times10^6$~yr. After this time, the fractional abundances start to decrease because molecules freeze out onto dust grains due to the low temperature ($T\simeq12$~K) and high density ($n\simeq10^{5}-10^{6}$~\cmt) achieved in these cores. On the other hand, the ion \htp\ keeps more or less a constant abundance until $t\simeq2.5\times10^6$~yr, and then its abundance rises moderately.

The primary formation route for \nth\ is through the reaction between H and N$_2$, which also produces 
H$_2$. \nth\ can then react with many atoms and molecules such as C, O, H$_2$O or CO to form N$_2$. 
Hence it is therefore clear that the abundance in the gas phase of these removal agents will 
be critical in determining the abundance of \nth\ present. At high densitites, but before the effects of freeze-out dominate, all the molecular species that are important in the removal of \nth\ and \nh\  will be high. In particular, for both models the main formation route for \nth\ is initially \hh\ + N$_{2}^{+}$ $\to$ \nth\ + H, and then the chemical reaction \htp\ + \nn $\to$ \nth\ + H becomes important, while \nh\ is mainly formed via dissociative recombination of NH$_{4}^{+}$. The latter is formed via consecutive hydrogenation of N$^{+}$ (via reactions with molecular hydrogen).

During phase~I  the \nh/\nth\ abundance ratio behaves in the same way for both the central core and the western/eastern cores models, and only during the latest time steps it depends on the adopted density ($n\simeq10^5$~\cmt\ for the western/eastern cores and $n\simeq10^6$~\cmt\ for the central core). In both cases, the \nh/\nth\ ratio starts from low values and it reaches values around 1000 and even higher for $t>2.5\times10^6$~yr because CO depletion favors the formation of \nh\ against \nth\ in core centers at high densities. This suggests that  long-lived starless cores  are associated with high values of the \nh/\nth\ abundance ratio.

%(EXPLAIN WHY NH3/N2H+ IS SO HIGH AT THE END OF THIS PHASE, NOTE THAT THE HIGEST VALUES OF THE RATIO ARE REACHED WHEN NH3 AND N2H+ %STARTS TO FREEZE OUT.)}

% In the following sections we present the main results of our chemical model obtained for phase~II calculation, after the protostar is born.}

%\begin{figure}[!ht]
%\begin{tabular}{c}
%    \epsfig{file=plots/ratio_temp.ps,scale=0.38, angle=0}\\
%    \end{tabular}
%     \caption{\nh/\nth\ abundance ratio as a function of temperature for the central core (model~D).} 
%\label{ratiotemp}
%\end{figure}

\subsubsection{Fractional abundances and the \nh/\nth\ abundance ratio during phase~II}

Figure~\ref{modelfigs}~(right panels) shows the fractional abundances (with respect to hydrogen nuclei) as a function of time of CO, N$_2$, H$_{3}^{+}$, \nh, \nth, and \hcop\  derived from our chemical model during phase~II for the central core (top panel) and the western/eastern cores (middle panel). The fractional abundances in the central core clearly have a different behavior compared to that of the western and eastern cores as a consequence of the different density, and especially the different temperature reached at the end of phase~II. In the central core, the fractional abundance of CO, N$_2$, \nth, and \hcop\ is constant during the initial stages, and
once the temperature increases, their fractional abundances rise significantly. On the other hand, the fractional abundance of the molecular ion \htp\ decreases with time. This is of course due to the fact that \htp\ is highly reactive at high densities. The abrupt increase of the fractional abundance takes place at the same time for all molecules, around $t\simeq6.1\times10^3$~yr (corresponding to a temperature of $T\simeq21$~K). This time corresponds to the time when a fraction of weakly bound species evaporates from the grain mantle (see Table 1 in Viti et al. 2004). The fractional abundance of \nth\
reaches its maximum value  $4.6\times10^{-11}$ (corresponding to a \nth\ column density of $\sim1.8\times10^{12}$~\cmd) at $t\simeq1.7\times10^4$~yr. The rapid destruction of \nth\ (by the chemical reaction \nth\ + CO $\to$ \hcop\ + N$_2$) occurs at $t\simeq2.3\times10^4$~yr, with a variation of the fractional abundance of one oder of magnitude, being the \nth\ column density around $\sim8\times10^{11}$~\cmd, consistent with the value estimated from the observations. On the other hand,  \hcop\  fractional abundance is constant at $\sim6\times10^{-10}$. Assuming [\hcop/\htcop]=40, we estimated an \htcop\ column density of $\sim1.3\times10^{12}$~\cmd, in good agreement with the estimation from the observational data (see Table~\ref{coldens}). This chemical behaviour could be explained in terms of CO desorption from grain mantles, which causes a substantial destruction of \nth\ favoring the formation of \hcop, being this reaction the main removal mechanism of \nth\ for
`standard' CO abundances of [CO/\hh]$\simeq10^{-4}$ \citep{jorgenssen2004,lee2004}. In fact, as can be seen in Fig.~\ref{modelfigs}~(top panel) the CO fractional abundance is $\sim10^{-4}$, high enough to lead a significant destruction of \nth. 

Contrary to the case of \nth\ and \hcop, the
increase of the \nh\ fractional abundance is slow, which indicates that gas phase chemistry dominates over pure evaporation. 
This is simply a consequence of the fact that most of the ammonia is still locked in water ice; in fact, as shown by the experiments by Collings et al. (2004), ammonia is 
released back into the gas phase only if temperatures of $\sim$ 100-120 K are reached (Viti et al. 2004). The maximum value, of 4.1$\times10^{-8}$ (or N(\nh)$\simeq3.2\times10^{15}$~\cmd), is reached at $t\simeq1.3\times10^5$~yr, and then it drops until it reaches a constant abundance of
1$\times10^{-9}$ (or a column density of $\sim8.2\times10^{13}$~\cmd).  This behavior, \ie\ the decrease in the \nh\ fractional abundance, takes place through reactions of the \nh\ molecule with the ions C$^+$ and \hcop. In this situation, the C$^+$ fractional abundance increases at late stages, 
while the fractional abundance of \hcop\ is more or less constant, which is consistent with the fact that \hcop\ is produced from the destruction of \nth\ (see above) and destroyed through reactions with \nh.

%SV: GEMMA, ISN'T THIS SIMPLY TO DO WITH THE FACT THAT NH3 DOES NOT EVAPORATE AT THESE TEMPERATURES? IF SO i SUGGEST THE FOLLOWING TEXT: 

Regarding the western and eastern cores, which were modeled assuming the same core size ($\sim0.04$~pc) and same temperature (the maximum temperature is 25~K), we found that the fractional abundance of \nh\  increases moderately with time, while the fractional abundances of CO, \nn, \nth, and \hcop\  rise considerably during the initial stages due to desorption effects, similar to the case of the the central core model. At $t\simeq4\times10^3$~yr, \nth\ and \hcop\ have constant abundance, around $\sim2\times10^{-10}$ (N(\nth)$\simeq4\times10^{12}$~\cmd) and $\sim1.8\times10^{-9}$ (N(\hcop)$\simeq3.6\times10^{13}$~\cmd\ or N(\htcop)=9$\times10^{11}$~\cmd, assuming [\hcop/\htcop]=40) for \nth\ and \hcop, respectively, while the \nh\ fractional abundance increases, up to $\sim5.2\times10^{-8}$ (N(\nh)$\simeq1\times10^{15}$~\cmd) at $t\simeq1.3\times10^{5}$~yr, and then it remains roughly constant too. For these cores, the values obtained from the chemical modeling are in agreement (within a factor of 2 in the case of \nth, and a factor of 4 in the case of \nh) with the column densities reported from the observational data. Finally, the CO fractional abundance in the western/eastern cores is $\sim10^{-5}$, significantly lower than the CO fractional abundance of the central core, indicating that the relatively low CO abundance in the western/eastern cores does not lead to a substantial destruction of \nth. Therefore, the fraction of CO that will evaporate from grain mantles plays an important role in determining the fractional abundance of \nth, and hence the \nh/\nth\ abundance ratio.

In Table~\ref{tableres} we show the values of the \nh\ and \nth\ column densities, together with the \nh/\nth\ abundance ratio
for the two models that are in agreement with the observed values. In Fig.~\ref{modelfigs}~(bottom right panel) we present the \nh/\nth\ abundance ratio as a function of time for the central core and the western/eastern core obtained during phase~II. The \nh/\nth\ abundance ratio observed toward the central core, around $\sim$400--1000, can be reproduced by our chemical model for times $t\simeq10^4$~yr and $t=$(4.5--5.3)$\times10^5$~yr. For the time range $t\simeq10^4$--$4.5\times10^5$~yr the model produces a higher \nh/\nth\ abundance ratio, $\sim4000$, slightly above the observed values for the central core. However we adopted the longer age as it is more realistic. For the western/eastern cores, the \nh/\nth abundance ratio initially shows high values, then it decreases due to desorption effects and finally reaches a constant value $\sim200$ for typical ages of low-mass YSOs, at around $t\simeq10^{5}-10^{6}$~yr. In addition, in Fig.~\ref{modelfigs} (bottom panel) we also show the temperature as a function of time. For the central core,  when the temperature is low, the \nh/\nth\ ratio is high ($>10^4$). Around $T\simeq21$~K there is a clear drop of the \nh/\nth\ ratio due to desorption effects. As temperature increases, the \nh/\nth\ rises until it reaches a constant value, $\sim10^3$, at around $T\simeq45$~K. For the western and eastern cores the temperature varies from 12 to 25~K, producing small variations on the \nh/\nth\ ratio, and only for times in the range $10^3$--$10^4$~yr the \nh/\nth\ ratio changes significantly, similar to the case of the central core. It is worth noting that in the model for the central core the visual extinction is $A_{\mathrm{V}}\simeq40$, but typically the visual extinction in hot cores (\ie\ embedded in the central core) is around $\sim100$ or even higher. In order to evaluate the error in the \nh/\nth\ abundance introduced by this difference we performed an additional model with a higher density, which gives a visual extinction of $A_{\mathrm{v}}\simeq100$. In this situation, the \nh/\nth\ ratio is affected by a difference of
$\sim3$--25~\%.We also note that the abrupt drop of the \nh/\nth\ ratio produced at early ages ($t\simeq5\times10^3$~yr) in both the central core and western/eastern cores is due to desorption effects of some molecules.

%However we adopted the longer age as it is more realistic. 
%SV: GEMMA, I AM PUZZLED BY THE DIFFERENCES YOU GET IF YOU INCREASE FROM 40 TO 100 MAG IN AV: WHAT ARE THESE DIFFERENCES DUE TO? ARE THEY %'LEFT OVER' DIFFERENCES FROM PHASE I? THE ONLY DIFFERENCE IT SHOULD MAKE IS IN PHASE I BEFORE IT REACHES THESE FINAL AV I.E WHEN RADIATION %CAN STILL PENETRATE....} 

Finally, we explored the fractional abundances of other molecular species and compared them with observations when available or made some predictions for future observations. We found that the fractional abundance of \chtcn\ obtained from the chemical model is in the range $\sim8\times10^{-8}-10^{-7}$ at $t=(4.5-5.5)\times10^{5}$~yr , in agreement with the value reported by \citet{zhang2007}, which is in the range $(1-4)\times10^{-8}$. Additionally, our model predicts a CO abundance of $10^{-4}$ indicating that CO is desorbed from grain mantles due to the relatively high temperature (around $\sim70$~K) and the powerful molecular outflows detected in CO \citep{zhang2007}. We found that molecules like HC$_3$N, H$_2$CO, C$_2$H, and CS should be detectable toward the central core, since their fractional abundances are even higher than the \nh\ ones. Concerning the western and eastern cores, the CO fractional abundance, of $10^{-5}$, is one order of magnitude lower than that of the central core. For these cores, molecules such as CS, C$_2$H, and \chtcn\ are quite abundant  and would be detectable but  we would not expect to detect any H$_2$CO and HC$_3$N.

In summary, the models that can reproduce the observed values of the \nh/\nth\ abundance ratio toward the central core as well as in the western and eastern cores, were obtained with a high depletion at the end of phase~I. Hence, the differences between the central and the western/eastern cores seen in the \nh/\nth\ 
abundance ratio seem to be mainly due to density and temperature effects. 
In particular, as shown by the chemical modeling of the region, the temperature is a key parameter in determining the \nh/\nth\ abundance ratio, and hence the abundance of each molecule. As temperature rises the fraction of species that will evaporate increase, and consequently the chemistry evolves differently.

% \textbf{In particular, the fraction of CO that will evaporate from grain mantles plays an important role in the fractional abundance of \nth, (\ie\ in the \nh/\nth\ abundance ratio). }

\section{Discussion}

The results obtained from the \nth\ observations, together with \nh\ data from the literature \citep{zhang2002} show that there are three dense cores associated with the high-mass star-forming region AFGL\,5142. The central core, strong in \nh\ but almost devoid of \nth, harbors a cluster of massive stars in their making. On the other hand, the western and eastern cores are stronger in \nth\ than in \nh\ emission. In the following section we discuss the possible reasons for the observed variation of the  \nh/\nth abundance ratio among the cores found in AFGL\,5142.

%\subsection{The \nh/\nth\ behavior in AFGL\,5142}

The central dense core in AFGL\,5142 is actively forming a cluster of massive
stars, still in the accretion phase, which displays the typical
signposts of star formation (molecular outflow and maser emission). Concerning the western and eastern cores, the relatively high temperature, around 20~K, could be due to the presence of an embedded YSO and/or heating produced by the three molecular outflows  associated with the central core, which are strongly affecting the surrounding dense gas \citep{zhang2002,zhang2007}. Thus, the association of the western and eastern cores with an infrared source is not obvious. 

The \nh/\nth\ abundance ratio map presented in Fig.~\ref{fratio} shows
significant variations, with a ratio around $\sim$50--100 toward
the western and eastern cores, and a ratio, up to 1000, toward the central core, due to a significant drop in the
\nth\ abundance. The values found for the \nh/\nth\ abundance ratio in the
western and eastern cores are similar to those found in previous
studies for cores harboring YSOs, for both low- and high-mass star-forming regions \citep{caselli2002a,hotzel2004,palau2007,friesen2010}. These studies are all consistent with the fact that a high ratio ($\leq$300) seems to be associated with starless cores and a low ratio, around 60--90, is found toward the YSOs, suggesting an anticorrelation between the the \nh/\nth\ abundance ratio and the evolutionary stage. Thus,  the
high \nh/\nth\ ratio toward the central core does not follow the
anticorrelation between the \nh/\nth\ ratio and the evolutionary stage
of the core like in the studies mentioned above. The noticeable
increase of the \nh/\nth\ ratio in the central core suggests a strong
differentiation of the
 \nth\ abundance between the central core and the western/eastern
 cores. Below we briefly investigate the origin of such a
 differentiation.

\subsection*{UV radiation effects:} 

As suggested by \citet{qiu2008}, UV photons from IRAS\,05274+334 are
not likely to affect the dense gas, and the effects of the UV field from
the embedded protostar(s) do not seem to be a major issue in
determining the \nh\ and \nth\ abundance of the central core, since
the high visual extinction, $A_{V}\simeq40$~mag or higher, prevents UV photons
to penetrate the central core.

\subsection*{Excitation effects:} 

Even though we took into account
the effects of opacity and different \Tex\ in the calculation of the
column densities, we consider whether the different excitation
conditions among the dense cores may be the cause
for the lower abundance of \nth, \ie\ the high \nh/\nth\ ratio. 
The critical density of \nth\ is $\sim10^5$~\cmt\
\citep{jorgenssen2004a}, while the critical density of \nh\ is
$\sim10^4$~\cmt\ \citep{ho1977}. From the dust continuum emission we
estimated a density of $\sim10^6$~\cmt\ in the central core. In
addition, the density obtained from single-dish observations at
850~$\mu$m \citep{Jenness1995}, adopting a core size of $\sim26''$, is
$\sim4\times10^5$~\cmt. For these densities the transitions of both
\nh\ and \nth\ are thermalized, and thus excitation effects do not seem appropriate to
explain the \nth\ emission drop in the central core.

On the other hand, the hyperfine structure method in CLASS
assumes equal excitation temperature for all the hyperfine
components. \citet{daniel2006} show that this assumption does not hold for the case of high opacities due to radiative processes. Since in AFGL\,5142 we derived an opacity of
$\tau_{\mathrm{TOT}}\simeq$0.3--0.6 for the \nth\ molecule, the problem reported by \citet{daniel2006} is not likely affecting our result.

\subsection*{Physical and chemical effects:}  

The high values, up to 1000, of the \nh/\nth\ abundance ratio found in the central core of AFGL\,5142 can be reproduced by our chemical model  for high densities
($n\simeq10^6$~\cmt) and high temperaures ($T\simeq70$~K).  The chemical modeling performed in Sect.~5 indicates that both density and
temperature play an important role in determining the molecular
abundance of \nh\ and \nth\ , and hence their ratio. The central
core has a higher temperature and density than the western and eastern
cores. Thus, a different chemistry can develop due to CO evaporation 
from the grain
mantles. 
The CO desorption
in the central core of AFGL\,5142 leads to the destruction of \nth, and
hence the \nh/\nth\ abundance ratio increases considerably relative to
the value found in the western and eastern cores. This is supported by
the fact that \tco\ is not frozen out in the central core, whereas it is faintly detected in the western and eastern cores
\citep{zhang2007}. As pointed out in Sect~5.2.2, \citet{collings2004} find experimental evidences that the desorption of \nh\ from grain mantles takes place at a temperature of $\sim120$~K. Therefore, since in our model we assumed a maximum temperature of 70~K, the high \nh/\nth\ abundance ratio in the central core is mainly a consequence of the destruction of \nth\ by CO rather than an enhancement of \nh\ in the gas phase. The disappearance of \nth\ from the gas phase has
been reported by several authors, both in low- and high-mass
star-forming regions. In the low-mass regime, there are several
reports of central \nth\ depletion in Class~0/Class~I protostars (\eg\
IRAM\,04191+1522: \citealt{belloche2004}; VLA\,1623:
\citealt{diFrancesco2004}; L483~mm: \citealt{jorgenssen2004a};
Barnard~1c: \citealt{Matthews2006}; BHR~71\,IRS1:
\citealt{chen2008}). Regarding the high-mass regime,
 \citet{Pirogov2003,Pirogov2007} observed with a single-dish telescope a sample of dense cores associated with massive stars and star clusters containing IRAS point sources, and find that, for most of the sources, there is a decrease of the \nth \ abundance toward the dust column density peak. Furthermore,  interferometric observations of \nth\ toward the high-mass star-forming regions IRAS\,23033+5951  \citep{Reid2008} also reveal a significant destruction of \nth\ . Therefore, the high ratio measured toward the central core of AFGL\,5142 seems to be due to the rapid destruction of \nth\ when CO is released from grain mantles as seems to be the case for the other regions cited above. 
  
%However, it is worth noting that a core with a lower temperature, like the western and eastern cores, where T is around $\sim$20--25~K, may also be associated with a high %\nh/\nth\ abundance ratio when the density is $n\simeq10^6$~\cmt. Therefore, both temperature and density may play an in important role in determining the \nh/\nth\ %abundance ratio. 

%SV: I WOULD REMVOE THE NEXT SENTENCE SINCE YOU DID NOT INCLUDE SHOCKS IN THE MODEL
%Furthermore, shocks in molecular outflows returns CO to the
%gas phase causing the depletion of \nth\ through ion-molecule
%reactions (see \citealt{bergin1998,vanDishoeck1999}). 
   
\subsection*{Evolutionary effects:} 

%Our chemical model shows that the high values of the \nh/\nth\
%abundance ratio in the central core of AFGL\,5142 are reached at
%$t\simeq$4.5--5.4$\times10^5$~yr for densities of $n\simeq10^6$~\cmt,
%while the values found in the western and eastern cores are compatible
%with $t\simeq10^{4}$--3$\times10^{6}$~yr. However, the case of
%IRAS\,20293+3952 clearly differs from the results found in AFGL\,5142,
%in which the \nth\ emission is stronger in regions containing YSOs
%than in starless cores \citep{palau2007}, suggesting that the YSOs in
%IRAS\,2023+3952 might be at an earlier stage of evolution than those
%associated with AFGL\,5142. Therefore, in addition to the chemical
%effects, the evolutionary stage of each core may also play an
%important role in the variations found in the \nh/\nth\ abundance
%ratio

Although in our chemical model we did not include molecular outflows, they may strongly affect the surrounding dense gas, and hence the \nh/\nth\ abundance ratio. 
As already  discussed by \citet{chen2008}, who propose three stages of the interaction between the \nth\ emission and molecular outflows, based on the morphology of the \nth\ dense gas and the jet/outflow emission. Here we suggests a qualitative picture of the evolution of the \nh/\nth\ abundance ratio. First, in starless cores with high densities most molecules are highly depleted, favoring the formation of \nh\ and \nth. It seems
that the CO depletion favors the formation of \nh\ 
against \nth\  in core centers at high densities. Thus, the \nh/\nth\ ratio reaches high values,
consistent with the results found in low-mass starless cores
\citep{caselli2002a,hotzel2004,friesen2010} as well as for starless cores nearby massive stars, like in the high-mass
star-forming regions IRAS\,20293+3952 \citep{palau2007}. The high \nh/\nth\ ratio in starless cores has been reproduced by our chemical model shown in Fig.\ref{modelfigs}~(left panels). After this phase, once the
star has formed,  the \nh/\nth\ ratio decreases, as found in the
low-mass regime \citep{caselli2002a,hotzel2004,friesen2010}, and in YSOs close to the massive star (\eg\ IRAS\,20293+3952: \citealt{palau2007}; IRAS\,00117+612: Busquet \et\ in prep.). In this
situation, CO is less depleted, and \nth\ molecules in the envelope are entrained by the molecular outflow, being the \nth\ emission elongated in the direction of the molecular outflow, as found
in IRAS\,20293+3952 \citep{palau2007} and IRAS\,00117+6412
\citep{palau2010}. Finally, the next evolutionary stage involves the release of
CO by powerful molecular outflows, which leads to the destruction of
\nth and forms large holes in the envelope. Then, the \nh/\nth\ abundance ratio raises considerably, like
in the central core of AFGL\,5142. 

Therefore, either molecular outflows and density/temperature effects seem to be key pieces in the determination of the  \nh/\nth\ abundance ratio.
In order to quantify the importance of the temperature in the \nh/\nth\ ratio, we compared the results obtained in AFGl\,5142 with the low-mass star-forming region L483. This region seems to be in the same stage of disruption of the envelope as AFGL5142, as the N2H+ shows two clumps on both sides of the YSO 
(see \citealt{jorgenssen2004a,fuller2000}), being then in the later evolutionary stage of the scenario proposed above.  However, the NH3
emission in L483 is also showing a hole close at the position of the YSO, suggesting that possibly the \nh/\nth\ abundance ratio is much lower than the
values found in AFGL\,5142, around 400--1000. This seems to indicate that the effects of molecular outflows alone cannot explain the large values derived in AFGL\,5142, although to draw a firm conclusion this should be modeled. Thus, the high temperature reached in the central core due to the presence of hot
cores seems to affect significantly the \nh/\nth abundance ratio, indicating that the \nh/\nth\ ratio behaves differently than in low-mass star-forming regions.

\section{Conclusions}

We used CARMA to observe the 3.2~mm continuum emission and the \nth\
molecular line emission toward the massive protostellar cluster
AFGL\,5142. The \nth\ dense gas emission, which mimics the morphology
of \nh, shows two main cores, one to the west (the western core) of the
dust condensation of $\sim0.08$~pc of size, and another core to the
east (the eastern core), of $\sim0.09$~pc. Interestingly, toward the
dust condensation, of M$\simeq23$~\mo, the \nth\ emission drops
significantly, whereas the compact \nh\ core (\ie\ the central core)
peaks at this position. The analysis of the \nth\ column density and
the \nh/\nth\ abundance ratio indicates a strong differentiation in
the \nth\ abundance. While we found low values, around $\sim$50--100, of the \nh/\nth\
abundance ratio associated with the western and eastern cores, the
\nh/\nth\ ratio rises significantly, up to 1000, toward the central
core, in which the formation of several massive stars is taking
place. We performed a chemical modeling of the region using the
time-dependent chemical model UCL\_CHEM. The chemical model shows that a high \nh/\nth\ abundance ratio can be reproduced in three different situations: \emph{i)} in long-lived starless cores, \emph{ii)} in high density cores with low temperature, $\sim$20--25~K, and containing YSOs, and \emph{iii)} in high density cores with YSOs and high temperature.
For the central core we found that such high
value of the \nh/\nth\ abundance ratio  could be
explained with a density of $\sim10^6$~\cmt\ and a maximum temperature
of 70~K, producing a substantial destruction of the \nth\ molecule by
CO, which is desorbed from the grains mantles due to the high
temperature and possibly the molecular outflows associated with the massive
YSOs embedded in the central core. On the other hand, to reproduce the
observed \nh/\nth\ abundance ratio toward the western and eastern cores, we used a lower density, of
$\sim10^5$~\cmt, and temperature (maximum temperature reached is
25~K). These results suggests that, in addition to the core evolution,
the physical properties of the core, such as density and temperature, play an
important role in determining the \nh/\nth\ abundance ratio. In conclusion, in AFGl\,5142 we found that the \nh/\nth\ abundance ratio behaves differently to the low-mass case mainly due to the high temperature reached in hot cores.

%\begin{enumerate}
%\item 
%\end{enumerate}

%Our main conclusions can be summarized as follows:

\begin{acknowledgements} 

We thank the anonymous referee for his/her valuable comments and suggestions. G. B. is grateful to Dr. L. Testi and Dr. T. R. Hunter for providing us the \hcop\ and \htcop\ data. G.~B., R.~E., A.~P., and
\'A.~S.-M. are supported by the Spanish MEC grant AYA2005-08523-C03, and the MICINN grant AYA2008-06189-C03 (co-funded with FEDER funds). A.~P. is supported by a JAE-Doc CSIC fellowship co-funded with the 
European Social Fund. This research has been partially funded by Spanish 
MICINN under the ESP2007-65475-C02-02 and Consolider-CSD2006-00070 grants. Q. Z. acknowledges the support from 
the Smithsonian
Institute USS Endowment funds.

%This publication makes use of the data products
%from the Two Micron All Sky Survey, which is a joint project of the University of Massachusetts and the Infrared Processing and Analysis
%Center/California Institute of Technology, funded by the National Aeronautics and Space Administration (NASA) and the National Science
%Foundation. 

\end{acknowledgements}

%\begin{figure*}[t]
%\begin{center}
%\begin{tabular}{c}
%    \epsfig{file=plots/phaseI_nitrogenchem.eps,scale=0.5}\\
%    \end{tabular}
%     \caption{\textbf{Only for the referee. Plot not included in the paper.}}
%\label{f3mmcont}
%\end{center}
%\end{figure*}

\end{document}